\documentclass[12pt,aps,prd,preprint, tightenlines,superscriptaddress,amsmath,amssymb,nofootinbib]{revtex4-1} 

\usepackage[letterpaper,dvips,hmargin={1in,1in},vmargin={1in,1in}]{geometry}

\RequirePackage[colorlinks=true
,urlcolor=blue
,anchorcolor=blue
,citecolor=blue
,filecolor=blue
,linkcolor=blue
,menucolor=blue
,pagecolor=blue
,linktocpage=true
,pdfproducer=medialab
,pdfa=true
]{hyperref}

\allowdisplaybreaks

\usepackage{amsmath,amssymb,amsthm,amsfonts}
\usepackage{graphicx}
\usepackage{xspace}
\usepackage{subfigure}
\usepackage{dcolumn}	
\usepackage{hyperref}      	
\usepackage{bm}		
\usepackage{epsfig}
\usepackage{epstopdf}
\usepackage{setspace}
\usepackage[usenames, dvipsnames]{color}
\usepackage{slashed}
\usepackage{comment}
\usepackage{enumitem}
\usepackage{siunitx}
\usepackage{array,multirow}
\usepackage{feynmp}
\usepackage{datetime}
\usepackage{enumitem}
\usepackage{titlesec, titletoc}
\titleformat*{\section}{\boldmath\bfseries}
\titleformat*{\subsection}{\boldmath\bfseries}
\setlist[description]{leftmargin=0.4cm}

\makeatletter
\def\endfmffile{%
	\fmfcmd{\p@rcent\space the end.^^J%
		end.^^J%
		endinput;}%
	\if@fmfio
	\immediate\closeout\@outfmf
	\fi
	\ifnum\pdfshellescape>\z@
	\immediate\write18{mpost \thefmffile}%
	\fi}
\makeatother

\newcommand{\PRE}[1]{{#1}} 

\newcommand{\be}{\begin{equation}\begin{aligned}}
\newcommand{\ee}{\end{aligned}\end{equation}}
\newcommand{\beq}{\begin{equation}}
\newcommand{\eeq}{\end{equation}}
\newcommand{\beqa}{\begin{eqnarray}}
\newcommand{\eeqa}{\end{eqnarray}}

\newcommand{\ifb}{\text{fb}^{-1}}
\newcommand{\iab}{\text{ab}^{-1}}

\newcommand{\mev}{\text{MeV}}
\newcommand{\gev}{\text{GeV}}
\newcommand{\tev}{\text{TeV}}

\newcommand{\cm}{\text{cm}}
\newcommand{\m}{\text{m}}

\newcommand{\mrad}{\text{mrad}}

\newcommand{\g}{\text{g}}

\newcommand{\ns}{\text{ns}}

\renewcommand{\eqref}[1]{Eq.~(\ref{eq:#1})}
\newcommand{\eqsref}[2]{Eqs.~(\ref{eq:#1}) and (\ref{eq:#2})}
\newcommand{\secref}[1]{Sec.~\ref{sec:#1}}
\newcommand{\secsref}[2]{Secs.~\ref{sec:#1} and \ref{sec:#2}}

\newcommand{\figref}[1]{Fig.~\ref{fig:#1}}

\newcommand{\tableref}[1]{Table~\ref{table:#1}}

\newcommand{\appref}[1]{Appendix~\ref{sec:#1}}

\RequirePackage[normalem]{ulem}

\newcommand{\flare}{FLArE\xspace}
\newcommand{\flareten}{FLArE--10\xspace}
\newcommand{\flarehundred}{FLArE--100\xspace}

\begin{document}
 
\count\footins = 1000
 
\preprint{PITT-PACC-2101}
\preprint{UCI-TR-2021-01}

\title{ 
\PRE{\vspace*{0.3in} }
{\large Detecting Dark Matter with Far-Forward Emulsion \\
and Liquid Argon Detectors at the LHC } 
\PRE{\vspace*{0.4in} }
}

\author{Brian~Batell}
\email{batell@pitt.edu}
\affiliation{Pittsburgh Particle Physics, Astrophysics, and Cosmology Center, \\ Department of Physics and Astronomy, University of Pittsburgh, Pittsburgh, PA 15217, USA
\PRE{\vspace*{0.1in}}
}

\author{Jonathan~L.~Feng}
\email{jlf@uci.edu}
\affiliation{Department of Physics and Astronomy,  University of California, Irvine, CA 92697-4575, USA
\PRE{\vspace*{0.1in}}
}

\author{Sebastian Trojanowski\PRE{\vspace*{0.2in}}
}
\email{strojanowski@camk.edu.pl}
\affiliation{Astrocent, Nicolaus Copernicus Astronomical Center Polish Academy of Sciences, ul.~Bartycka 18, 00-716 Warsaw, Poland}
\affiliation{Consortium for Fundamental Physics, School of Mathematics and  Statistics, University of Sheffield, Hounsfield Road, Sheffield, S3 7RH, UK}
\affiliation{National Centre for Nuclear Research, Pasteura 7, 02-093 Warsaw, Poland
\PRE{\vspace*{0.3in}
}
}

\begin{abstract}
\PRE{\vspace*{0.2in}}
 New light particles may be produced in large numbers in the far-forward region at the LHC and then decay to dark matter, which can be detected through its scattering in far-forward experiments.  We consider the example of invisibly-decaying dark photons, which decay to dark matter through $A' \to \chi \chi$.  The dark matter may then be detected through its scattering off electrons $\chi e^- \to \chi e^-$.  We consider the discovery potential of detectors placed on the beam collision axis 480 m from the ATLAS interaction point, including an emulsion detector (FASER$\nu$2) and, for the first time, a Forward Liquid Argon Experiment (\flare).  For each of these detector technologies, we devise cuts that effectively separate the single $e^-$ signal from the leading neutrino- and muon-induced backgrounds.  We find that 10- to 100-tonne detectors may detect hundreds to thousands of dark matter events in the HL-LHC era and will sensitively probe the thermal relic region of parameter space.  These results motivate the construction of far-forward emulsion and liquid argon detectors at the LHC, as well as a suitable location to accommodate them, such as the proposed Forward Physics Facility.
\end{abstract}


\maketitle
\tableofcontents


\clearpage

\section{Introduction}
\label{sec:introduction}

Unraveling the nature of dark matter (DM) is one of the top priorities in particle physics and cosmology today. As the search for traditional DM candidates, such as weakly interacting massive particles (WIMPs) and axions, resolutely marches forward, physicists are aggressively exploring new frontiers in the vast DM landscape. One particularly compelling idea is that DM is part of a light hidden sector, coupled to the Standard Model (SM) through a light mediator particle. The observed DM relic abundance can be obtained through simple thermal freeze-out in these scenarios~\cite{Boehm:2003hm,Pospelov:2007mp,Feng:2008ya}. This extends the standard WIMP production mechanism to dark matter masses below the Lee-Weinberg bound~\cite{Lee:1977ua} and provides predictive targets in DM model parameter space in the MeV to GeV mass range~\cite{Izaguirre:2015yja,Krnjaic:2015mbs,Batell:2017cmf}. Light DM models of this kind lead to a rich phenomenology, requiring new experimental and observational search strategies going beyond the traditional methods used to search for WIMPs~\cite{Alexander:2016aln,Battaglieri:2017aum,Beacham:2019nyx}.

A promising avenue for light weakly interacting particle searches is to harness the large and energetic far-forward flux of particles produced in proton-proton collisions at the Large Hadron Collider (LHC). In particular, the FASER detector~\cite{Feng:2017uoz,Ariga:2018zuc,Ariga:2018pin}, situated 480 m downstream of the ATLAS interaction point (IP), will efficiently search for light long-lived particles emerging from this forward flux and decaying visibly to SM particles~\cite{Feng:2017uoz,Feng:2017vli,Batell:2017kty,Helo:2018qej,Kling:2018wct,Feng:2018pew,Dercks:2018eua,Ariga:2018uku,Kling:2020mch}. Furthermore, FASER$\nu$ will initiate studies of collider-produced neutrinos and measure TeV-energy neutrino interactions in a controlled laboratory setting for the first time~\cite{Abreu:2019yak,Abreu:2020ddv,Ismail:2020yqc}.  FASER and FASER$\nu$ are currently being constructed and installed to collect data during LHC Run~3, while a second stage of these experiments with much larger detectors is envisioned for the high-luminosity Large Hadron Collider (HL-LHC) era. These larger detectors are unlikely to fit within the existing tunnel infrastructure, but a dedicated Forward Physics Facility (FPF)~\cite{SnowmassFPF}, which would house these and other experiments to carry out a variety of novel SM measurements and beyond the SM (BSM) searches, is currently under study~\cite{FPFKickoffMeeting}.

In this paper we investigate the prospects to search for light MeV- to GeV-scale hidden sector DM in the far-forward region of the LHC. The basic detection strategy is very simple. Light mediator particles may be copiously produced in the forward region of LHC $pp$ collisions and promptly decay to light DM particles. The DM particles then travel roughly 480 m into a detector and scatter off electrons.  A similar search strategy, but for proton fixed target experiments, has been investigated~\cite{Batell:2009di,deNiverville:2011it,deNiverville:2012ij,Batell:2014yra,Dobrescu:2014ita,Kahn:2014sra,Coloma:2015pih,deNiverville:2015mwa,deNiverville:2016rqh,DeRomeri:2019kic,Dutta:2019nbn,Dutta:2020vop} and successfully carried out by the MiniBooNE-DM Collaboration~\cite{Aguilar-Arevalo:2017mqx,Aguilar-Arevalo:2018wea}. 
As we will demonstrate, despite the simplicity of the signal signature, the single scattered electron has kinematic characteristics that allow it to be distinguished from all neutrino-induced and other SM backgrounds. 
This stems from the fact that the DM-electron scattering is mediated by a light force carrier and so favors $\cal{O}(\gev)$ electron recoil energies, while the background from neutrino-electron scattering is mediated by heavy electroweak bosons and so is spread over a broad range of energies and peaks in the several hundred GeV range.  Note that DM scattering off nuclei is also potentially a promising signal, but one we will not investigate here.

Although there are many different light DM scenarios that can be considered, we will focus here on the simple and well-motivated scenario in which the mediator is a kinetically-mixed dark photon~\cite{Holdom:1985ag}. In this case, both the DM production and subsequent DM-electron scattering is mediated by the dark photon. We will consider two potential detector technologies: (1) a 10-tonne-scale  emulsion detector, FASER$\nu$2, which is essentially an upgraded and enlarged version of the FASER$\nu$ experiment, and, for the first time, (2) a Forward Liquid Argon Experiment, which we denote by the acronym FLArE, composed of a 10- or 100-tonne-scale liquid argon time projection chamber (LArTPC) of the type being employed in several modern neutrino experiments. We will show that these detector types offer the potential to detect hundreds to thousands of DM events and to discover DM in a large region of parameter space in which the correct DM abundance is obtained through thermal freeze-out. 

The possibility of detecting DM in the forward region has been discussed for the SND@LHC experiment~\cite{Ahdida:2020evc}, an 850 kg emulsion detector proposed to be placed 480 m from the ATLAS IP just off the beam collision axis during Run~3. For the simple model considered and the luminosity expected at Run~3, no parameter space beyond current bounds could be probed~\cite{Ahdida:2020evc}, but, of course, other models could be considered.  In this study, rather than consider such models, we focus on two minimal models, but explore the potential of larger detectors placed on-axis during HL-LHC running.  In addition, we discuss muon-induced backgrounds in detail and, as noted above, analyze the physics potential of far-forward LArTPC detectors for the first time. 

This paper is organized as follows. In \secref{model} we describe the dark photon-mediated DM models to be studied in this work. In \secref{Detectors} we discuss the basic detector designs we will consider, and in \secref{signal} we provide an overview of our modeling of the DM signal. Our estimates of the neutrino- and muon-induced backgrounds are detailed in \secsref{nubackgrounds}{mubackgrounds}, respectively. We present the results of the analysis in \secref{results} and our conclusions and outlook in \secref{conclusions}.

\section{Light Dark Matter with a Dark Photon Mediator}
\label{sec:model}

In this section we introduce two simple, predictive benchmark models of  sub-GeV DM that can be explored in the far-forward region at the LHC. The models we study are based on a broken $U(1)_D$ gauge symmetry, with a massive dark photon $A'_\mu$ serving as a mediator between the SM and DM $\chi$. The dark photon Lagrangian is given by
\be
\mathcal{L} \supset -\frac{1}{4} F'_{\mu\nu}F^{'\mu\nu} +\frac{1}{2} m_{A'}^2 A'_\mu A^{'\mu} + \frac{\epsilon}{2 \cos\theta_W} F'_{\mu\nu}B^{\mu\nu} \ ,
\ee
where $m_{A'}$ is the dark photon mass, $\epsilon$ is the kinetic mixing parameter, $\theta_W$ is the weak mixing angle, and $B_{\mu\nu}$ is the hypercharge field strength. In the regime $m_{A'} \ll m_Z$ of interest here, the $A'$ primarily mixes with the SM photon, leading to a coupling of the dark photon to the electromagnetic current with strength suppressed by $\epsilon$. In the physical basis, the dominant interactions of the dark photon are then given by
\be
\mathcal{L} \supset A'_\mu (\epsilon \, e \, J_{EM}^\mu + g_D \, J_D^\mu) \ ,
\ee
where $J_{EM}^\mu$ is the SM electromagnetic current, $J_D^\mu$ is the $U(1)_D$ current, and $g_D \equiv \sqrt{4 \pi \alpha_D}$ is the $U(1)_D$ gauge coupling. 

It remains to specify the precise nature of the DM particle $\chi$. In this work, we will study two cases: (1) complex scalar DM and (2) Majorana fermion DM, with Lagrangians
\be
\mathcal{L} \supset 
\begin{cases}
\displaystyle{\ |\partial_\mu \chi|^2 - m_\chi^2 |\chi|^2} \quad \text{(complex scalar DM)}  \\
\vspace{-10pt}\\
\displaystyle{\ \frac{1}{2} \overline \chi i \gamma^\mu \partial_\mu \chi 
-\frac{1}{2} m_\chi \overline \chi \chi} \quad \text{(Majorana fermion DM)} \ ,
\end{cases}
\ee
where $m_\chi$ is the DM mass. The $U(1)_D$ current in each model is
\be
\label{eq:JD}
J_D^\mu = 
\begin{cases}
\displaystyle{\ i \chi^* \overset{\text{\footnotesize$\leftrightarrow$}}{\partial_\mu} \chi} \quad \text{(complex scalar DM)}  \\
\vspace{-10pt}\\
\displaystyle{\ \frac{1}{2} \overline \chi \gamma^\mu \gamma^5 \chi} \quad \text{(Majorana fermion DM)} \ .
\end{cases}
\ee

In both models the correct DM relic abundance can be achieved through thermal freeze-out. In particular, in the regime $m_{A'} > 2 m_\chi$, DM annihilates directly to SM fermions through $s$-channel dark photon exchange, $\chi \chi \to A^{'(*)} \to f \bar f$. In the limit $m_{A'} \gg m_\chi$, the annihilation cross section has the same parametric form in both models,
\be
\label{eq:sigv}
\sigma v \propto \alpha \, v^2 \, \frac{\epsilon^2 \, \alpha_D \, m_\chi^2}{m_{A'}^4}  = \alpha \, v^2 \, \frac{y}{m_\chi^2} \ ,
\ee
where $\alpha$ is the SM electromagnetic fine structure constant, and we have introduced the dimensionless parameter $y \equiv \epsilon^2 \alpha_D (m_\chi/m_{A'})^4$, following Ref.~\cite{Izaguirre:2015yja}.  A notable feature of the annihilation cross section of \eqref{sigv} is the velocity suppression characteristic of $P$-wave annihilation. These models therefore evade in a simple way the otherwise stringent constraints from DM annihilation at the epoch of recombination, which can lead to distortions in the cosmic microwave background temperature anisotropies~\cite{Ade:2015xua,Slatyer:2009yq}.  

Regions of parameter space where the correct DM relic density is obtained by thermal freeze-out are important targets for experimental searches. These can be conveniently visualized in the $(m_\chi, y)$ plane and compared with existing experimental bounds and future sensitivity projections. 
We will use the results of Ref.~\cite{Berlin:2018bsc} for the DM relic abundance predictions, derived by evolving the full Boltzmann equation including exact thermal averaging of the annihilation cross section. 
To present thermal targets and experimental sensitivities in the $(m_\chi, y)$ plane, we will adopt the common choices of $\alpha_D = 0.5$ and $m_{A'}  = 3 m_\chi$. These choices are fairly conservative, in the sense that they do not inflate the potential of probing thermal targets at particle experiments. However, it is important to keep in mind that the parametric scaling of the annihilation cross section in \eqref{sigv} is sharply violated in the resonance region, where $m_{A'} - 2 m_\chi \ll m_{A'}$. In this region, the annihilation rate is resonantly enhanced, and so smaller couplings are required to achieve the correct relic density, making it more challenging to experimentally probe thermal targets~\cite{Feng:2017drg,Berlin:2020uwy,Bernreuther:2020koj}.

Although the particle nature and couplings of complex scalar and Majorana fermion DM models are distinct (see Eq.~(\ref{eq:JD})), the two models yield quantitatively similar predictions for reaction rates involving relativistic energies. This includes DM produced in the hot early universe, as well as DM production at accelerators and its subsequent re-scattering in downstream detectors. On the other hand, for non-relativistic processes, such as halo DM scattering in direct detection experiments, the two models lead to dramatic quantitative differences in event rates. For Majorana fermion DM, direct detection rates are suppressed by many orders of magnitude due to the inherent momentum dependence of the scattering, and DM direct detection is challenging for all existing and proposed experiments.  (See, however, the recent study of Ref.~\cite{Kahn:2020fef}.)  In contrast, for complex scalar DM that scatters elastically, large event rates are expected in next generation sub-GeV direct detection experiments, complementing accelerator probes, including far-forward production at the LHC. We note, however, that even in the scalar DM case, one can introduce a mass splitting so that the scattering proceeds inelastically~\cite{TuckerSmith:2001hy}. Provided the fractional mass splitting is in the range ${\cal O}(10^{-6} - 10^{-1})$, this effectively suppresses scattering rates in direct detection experiments while leaving the cosmology and accelerator probes unaltered. These features will be clearly illustrated when we present our results in \secref{results}. 

We note that other viable models with a dark photon mediator can be constructed; see e.g., Ref.~\cite{Berlin:2018bsc} for a detailed discussion of model variations. In particular, certain models also lead to visible long-lived particle signatures, e.g., the visible decay of the dark photon mediator for the case $m_\chi > m_{A'}$, or the decay of the excited DM state $\chi_2 \to \chi_1 \, e^+ e^-$ in inelastic scenarios. Such long-lived particle signatures can also be explored at the LHC with the FASER experiment~\cite{Feng:2017uoz,Berlin:2018jbm,Jodlowski:2019ycu}.

\section{Detectors}
\label{sec:Detectors}

As already mentioned in \secref{introduction}, our analysis will focus on two distinct detector designs sensitive to the signal of DM-electron scattering. Both types of detectors have been successfully employed in past experiments and will be used in future searches. Here, we briefly present the basic details of the detectors we will consider. 

\subsection{Emulsion Detector: FASER$\nu$2}

We first consider an emulsion detector similar to, but larger than, the FASER$\nu$ detector currently under construction~\cite{Abreu:2019yak,Abreu:2020ddv}. FASER$\nu$ is a 1.1-tonne neutrino detector, composed of tungsten sheets for the target material, interleaved with emulsion films capable of detecting charged tracks with high spatial resolution.  This detector design will be tested in the far-forward region of the LHC during Run~3.

For the HL-LHC era, a larger 10-tonne scale emulsion detector, referred to as FASER$\nu$2, is currently envisioned~\cite{SnowmassFASERnu2}.  For our study, we will assume a rectangular detector with location and size given by
\begin{equation}
\text{FASER$\nu$2}: \ L = 480~\m \, ,\ \Delta = 2~\m \,  , \ S_T = (0.5~\m \times 0.5~\m) \, ,
\label{eq:FASERnu2}
\end{equation}
where $L$ is the distance from the IP to the front of the detector, and $\Delta$ and $S_T$ are the longitudinal and transverse dimensions of the tungsten target. We assume that the detector is centered on the beam collision axis.  

Tungsten-emulsion detectors have significant virtues.  They are remarkably compact, as a result of tungsten's high density of $19.3~\g/\cm^3$, and the excellent spatial and angular resolutions of emulsion detectors make them very precise tools for reconstructing individual interaction vertices in the detector~\cite{Nakamura:2006xs}. In particular, a particle track spatial resolution down to $50~\textrm{nm}$ can be achieved, while, depending on the track length in the emulsion, the angular resolution can be much better than $1~\mrad$. The expected energy resolution for $\sim\gev$ electromagnetic showers is $30-40\%$ and improves for higher energy showers; see, e.g., the discussion in Ref.~\cite{Agafonova:2010dc} for showers initiated by two photons. We will assume that tracks down to momenta of 300 MeV can be detected, and that the emulsion is exchanged periodically so that the track density remains manageable and suitable high speed scanning facilities are available. 

The main disadvantage of emulsion detectors for this DM search is the lack of timing, which makes it difficult to reject muon-induced background, as we discuss in \secref{mubackgrounds}. To remedy this, it is necessary to augment the emulsion-tungsten detector with interleaved electronic tracker layers, which would provide event time information. A similar detector concept was successfully employed in the OPERA experiment~\cite{Acquafredda:2009zz}, and analogous designs have been proposed for SND@SHiP~\cite{Anelli:2015pba} and SND@LHC~\cite{Ahdida:2020evc}. The role of such tracker layers in mitigating potential muon-induced backgrounds will be discussed further in \secref{mubackgrounds}.

\subsection{Forward Liquid Argon Experiment: \flare} 

Although the emulsion detector technology allows for a compact detector with the excellent energy resolution and vertex reconstruction capabilities required for the DM search, other experimental approaches are possible, especially if more space is available in a future Forward Physics Facility during the HL-LHC era~\cite{SnowmassFPF}.  

Liquid argon has proven to be a very efficient active detector material and has been successfully used in numerous DM direct detection searches and neutrino experiments. In particular, LArTPCs provide precise spatial and calorimetric resolution, excellent particle identification capabilities, and detailed neutrino event reconstruction.  Well-known examples include the short-baseline neutrino program at Fermilab~\cite{Antonello:2015lea} and the far detector of the future Deep Underground Neutrino Experiment (DUNE)~\cite{Abi:2020evt}. Current and planned detectors of this type have masses as large as $\sim 10^4$ tonnes. LArTPC detectors can therefore offer both large event statistics and excellent background rejection capabilities.

With this motivation, we will investigate the feasibility of detecting DM in a LArTPC detector in the far-forward region of the LHC. We will refer to this proposal as the Forward Liquid Argon Experiment (\flare). Fittingly, as we discuss in detail below, the first indication of a DM signal event in \flare is the flare of scintillation light produced by the electron recoil, followed by the ionization electron signal.

Since liquid argon is about 14 times less dense than tungsten, a LArTPC detector must be much larger if a similar active material mass is desired. In the models considered here, the DM is produced in light dark photon decay and so is highly collimated.  At a distance of 480~m from the IP, the flux is largely confined to a region of a few tens of cm around the beam collision axis.  The optimal detector therefore has a cross sectional area of approximately this size. Note, however, that in other BSM scenarios in which DM particles are produced with broader angular distributions, for example, from the decay of heavier mesons, increasing the transverse size of the detector could also be beneficial.

Motivated by these considerations, we will consider two liquid argon detectors: 
\begin{eqnarray}
&& \text{\flareten (10 tonnes)}: \ \ \quad  L = 480~\m,\ \Delta = 7~\m,\ \ \, S_T = (1~\m \times 1~\m) \ ,
\label{eq:LAr10ton} \\
&& \text{\flarehundred (100 tonnes)}: \ \ L = 480~\m, \ \Delta = 30~\m,\ S_T = (1.6~\m \times 1.6~\m) \ ,
\label{eq:LAr100ton}
\end{eqnarray}
where, as above, $L$ is the distance from the IP to the front of the detector, $\Delta$ and $S_T$ are the longitudinal and transverse dimensions of the detector, and we assume that the detector is centered on the beam collision axis.  These detectors would require enlarging the available space in the far-forward region of the LHC, which, however, has already been envisioned in the proposed Forward Physics Facility~\cite{SnowmassFPF}. 

LArTPC detectors can detect very soft charged tracks down to momenta of 30 MeV, a significant improvement over emulsion detectors.  In addition, as noted above, emulsion detectors suffer for this analysis from the lack of timing, which makes it difficult to reject muon-induced backgrounds.  In contrast, the expected time resolution of LArTPC detectors is at the level of $\mathcal{O}(\textrm{ms})$ due to the finite drift time of ionization electrons in liquid argon~\cite{Acciarri:2016smi}. Further improvement can be achieved by using an additional light collection system in tandem with the TPC. Such a design has already been employed in the MicroBooNE experiment~\cite{Acciarri:2016smi}. This is used in the initial step of the background rejection procedure, which is based on the measurement of the light collected in consecutive $\mathcal{O}(10~\ns)$ long time ticks~\cite{MicroBooNE:2019knj}.\footnote{While MicroBooNE has photomultiplier tubes (PMTs) behind the wire planes, a future LArTPC could employ another PMT array on the other side of the detector, which could potentially significantly improve the event time resolution and location determination~\cite{Milind}.}  The combined event time and spatial information can be utilized to efficiently separate the DM signal from the muon-induced backgrounds, as we discuss in more detail in \secref{mubackgrounds}. 

We note that further redundancy could be provided by a dedicated muon-tagger positioned either in front of or behind the LArTPC. In a similar fashion to the MicroBooNE Cosmic Ray Tagger~\cite{Adams:2019bzt}, such a system could measure the crossing time and position of the passing muons, which could then be compared with similar information obtained with the TPC and light detection system.  Notably, the time and position information of the through-going muons can also be obtained with the use of other experiments placed in the Forward Physics Facility along the beam collision axis, e.g., by employing the tracking system of the proposed FASER 2 experiment. 

\section{Signal}
\label{sec:signal}

In the previous sections, we introduced the models and detectors we will consider.  Here we discuss in detail how we simulate the signal, the characteristics of the single electron signature, and suitable cuts to distinguish signal from background.

\subsection{Signal Simulation}

For the models discussed in \secref{model}, at the LHC, DM is primarily produced through the decays of on-shell dark photons. The parent dark photons with $m_{A^\prime}\alt 1~\gev$ are dominantly produced in rare decays of neutral pions and $\eta$ mesons, as well as through proton bremsstrahlung. 

In our modeling, we employ the \texttt{CRMC} simulation package~\cite{CRMC} and the \texttt{EPOS-LHC} Monte Carlo (MC) generator~\cite{Pierog:2013ria} to obtain the forward spectrum of light mesons produced in $14~\tev$ proton-proton collision energy. As a cross check, we have also used the meson spectra obtained with \texttt{Pythia\,8}~\cite{Sjostrand:2014zea} and found good agreement.  A MC simulation then generates the rare decays $\pi^0,\eta\to\gamma A^\prime$, the subsequent prompt dark photon decays $A^\prime\to\chi\chi$, the propagation of the DM particles to the detector, and the interactions of the DM particles in the detector, $\chi e^- \to\chi e^-$. To model $A'$ production from proton bremsstrahlung, $pp\to ppA^\prime$, we use the Fermi-Weizsacker-Williams approximation, following the discussion in Refs.~\cite{Blumlein:2013cua,Feng:2017uoz,deNiverville:2016rqh}. 

An additional flux of light DM particles in the forward direction is generated when high-energy photons and neutral hadrons hit the neutral particle absorber TAXN located about $130~\m$ away from the $pp$ IP.  These then generate electromagnetic (EM) showers and hadronic cascades, which can then produce DM particles in the TAXN. Similar recent discussions of DM production in electron- and proton-induced EM showers in the target for selected beam-dump and neutrino experiments can be found in Refs.~\cite{Marsicano:2018glj,Celentano:2020vtu}.

We have made preliminary estimates of the TAXN-produced DM flux for the FASER$\nu$2 detector discussed in \secref{Detectors}.  We find that the dominant source is from EM showers producing dark photons through resonant production, $e^+e^-\to A^\prime$, and from $A^\prime$ production through electron bremsstrahlung, but even these production modes are subdominant in our simulations. This is due to the relatively high electron recoil energies of interest for our signal regions, $E_e \agt 300~\mev$, and the small angular size of FASER$\nu$2, which covers only a small fraction of the total angular spread of the EM showers in the TAXN.  We note, however, that for the larger LArTPC detectors with a lower electron recoil energy threshold, DM production in the TAXN could play a non-negligible role, especially for low DM masses $m_\chi \alt 10~\mev$. We leave a detailed analysis of this effect for future studies.

\subsection{Signal Characteristics}

\begin{figure}[t]
\begin{center}
\includegraphics*[width=0.25\textwidth]{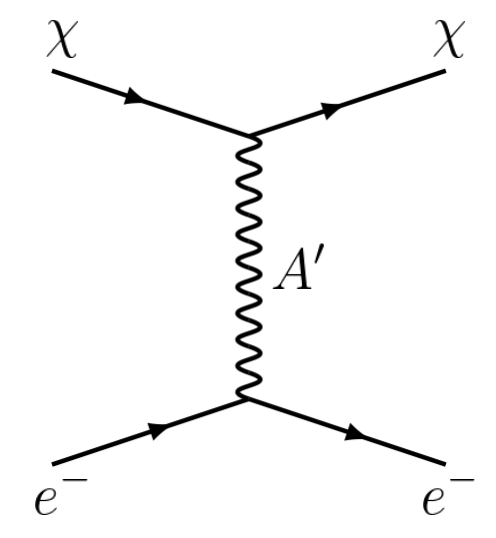}
\hfill
\end{center}
\vspace*{-0.3in}
\caption{The DM-electron elastic scattering signal process. }
\label{fig:DM-e}
\end{figure}

The signature of our interest is a single-electron-initiated EM shower in the detector that results from $\chi e^- \to \chi e^-$ scattering, as shown in \figref{DM-e}. For the complex scalar and Majorana fermion DM models, the differential scattering cross sections are
\begin{eqnarray}
\label{eq:complscat}
\hspace*{-0.3in} \frac{d\sigma}{d E_e} \! &=& \! 4 \pi \epsilon^2 \alpha \alpha_D \frac{2 m_e E^2 - (2 m_e E+ m_\chi^2)(E_e-m_e)}{(E^2-m_\chi^2)(m_{A'}^2 + 2 m_e E_e -2 m_e^2)^2 } \quad \textrm{(complex scalar DM)} \, , \\
\hspace*{-0.3in} \frac{d\sigma}{d E_e} \! &=& \! 4 \pi \epsilon^2 \alpha \alpha_D 
\frac{2 m_e (E^2 \! - \! m_\chi^2) \! + \! [ m_\chi^2 \! - \! m_e (2E \! - \! E_e \! + \! 2 m_e)](E_e \! - \! m_e)}{(E^2-m_\chi^2)(m_{A'}^2 + 2 m_e E_e -2 m_e^2)^2 } \quad  \textrm{(Majorana DM)} \, ,
\label{eq:Majoscat}
\end{eqnarray}
where $E_e$ is the electron recoil energy and $E$ is the energy of the incoming DM particle. For large DM energies $E\gg E_e,m_\chi$ and electron recoil energies satisfying $E_e\gg m_e$, the scattering cross sections in these two scenarios become very similar, and they are both well approximated by
\begin{flalign}
\label{eq:Approxscat}
\frac{d\sigma}{d E_e} & \approx \frac{8 \pi \, \epsilon^2 \, \alpha \, \alpha_D \, m_e }{(m_{A'}^2 + 2 m_e E_e)^2 } \qquad \textrm{(approximate formula for large $E, E_e$}) \ .
\end{flalign}
For further details regarding these differential cross sections, see \appref{scattering}.

\begin{figure}[t]
\begin{center}
\includegraphics*[width=0.49\textwidth]{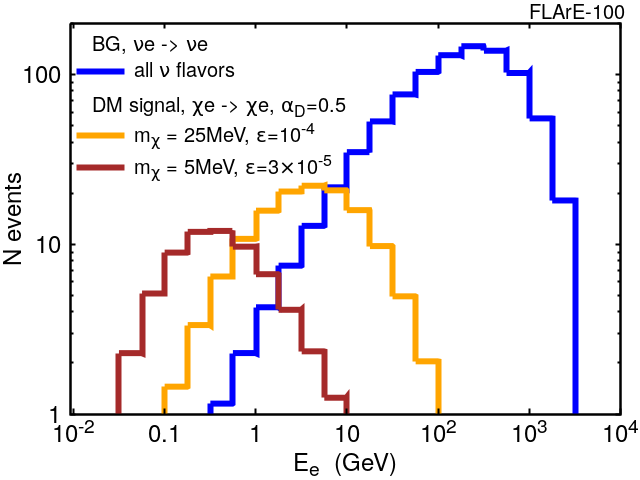}
\hfill
\end{center}
\vspace*{-0.3in}
\caption{The electron recoil energy spectra for the background neutrino-electron scattering events (blue histogram) and for the DM signal events. The latter have been obtained for two benchmark scenarios with the DM mass and the kinetic mixing parameter equal to $(m_{\chi},\epsilon)=(25~\mev,10^{-4})$ (yellow) and $(5~\mev,3\times 10^{-5})$ (red). In both the DM scenarios we set the dark photon mass to $m_{A^\prime} = 3\,m_\chi$ and the dark coupling constant to $\alpha_D=0.5$. The number of events in the bins is normalized to the \flarehundred detector and the HL-LHC phase with $3~\iab$ of integrated luminosity and $14~\tev$ proton-proton collision energy. } 
\label{fig:1Dspectra}
\end{figure}

Despite the suppression by the small kinetic mixing $\epsilon$, the presence of the small dark photon mediator mass in the denominator of \eqref{Approxscat} can increase the DM scattering rates to a level that is comparable to or larger than those of SM neutrinos.  This is especially relevant for small electron recoil energies.  We illustrate this in \figref{1Dspectra} by comparing the $E_e$ spectrum of the dominant neutrino-electron scattering background to the $E_e$ spectrum of the DM signal in \flarehundred. 
As can be seen, in the case of the neutrino-induced events, the electron recoil energy is peaked at high (TeV) energies and the number of events decreases for low values of $E_e$. This is a result of the parent neutrino energy spectrum, which is peaked at a few hundred $\gev$, and the increase of the neutrino scattering cross section with neutrino energy. In addition, given the relatively large mass of the mediator $W$ and $Z$ bosons, for a fixed incident neutrino energy $E_\nu$, the relevant differential cross section depends only mildly on $E_e$.

On the other hand, the low dark photon mediator mass yields a DM scattering cross section that is largely independent of the incident DM energy. Also, the differential cross section is peaked towards small energy transfer to the recoiled electron such that $2m_e E_e\lesssim m_{A^\prime}^2$, which minimizes the denominator in \eqref{Approxscat}. The signal event rate can therefore be greatly enhanced for detectors with low energy thresholds, which becomes increasingly important with the decreasing dark photon mass. This can also be seen in \figref{1Dspectra}, where, keeping in mind $m_{A'} = 3 m_\chi$, the electron recoil energy spectrum for the benchmark scenario with $m_\chi=5~\mev$ is shifted towards lower $E_e$ in comparison with the similar spectrum obtained for $m_\chi=25~\mev$. As we illustrate below in \secref{results}, this allows the LArTPC detectors to probe smaller couplings than the emulsion detector in the regime with $m_\chi\lesssim \mathcal{O}(10~\mev)$.

An additional handle to distinguish the signal from the neutrino-induced background is the electron recoil angle, $\theta_e$, which can be written as 
\begin{equation}
\cos{\theta_e} \approx \frac{E_{\chi/\nu} \, E_e-m_T(E_{\chi/\nu} - E_e)} {\sqrt{E_{\chi/\nu}^2-m_{\chi/\nu}^2} \sqrt{E_e^2-m_e^2}} \ ,
\end{equation}
where $E_{\chi/\nu}$ is the incident energy of the DM particle or neutrino, and $m_T$ is the target mass. The target mass is equal to $m_e$ for the signal and the $\nu e^- \to \nu e^-$ background, but it is the nucleon mass for neutrino-induced backgrounds from quasi-elastic processes, $\nu (p/n)\to e (n/p)$. For the typical case, where $E_{\chi}\gg m_\chi$ and $E_e\gg m_e$, $E_e\,\theta_e^2 \approx 2m_T(1-x)$, where $x = E_e/E_{\chi/\nu}$ is the energy transfer to the electron. 

As discussed above, for the signal, the scattering rate is greatly enhanced for low values of $E_e$ and $x$. This implies that $\theta_e$ is typically larger for the DM signal than for the $\nu e^- \to \nu e^-$ background.  For the DM signal, for $x \alt 0.1$ and $E_e\sim \mathcal{O}(10~\gev)$, the typical recoil angle is $\theta_e \approx \sqrt{2m_e/E_e} \sim 10~\mrad$. 

On the other hand, for neutrino scatterings off nucleons, the typical recoil angle $\theta_e$ is much larger than in the case of the DM signal, given the much increased target mass. This helps to differentiate between the DM signal and these background events, which may be further suppressed by requiring no additional charged tracks in the detector. These considerations are especially relevant for deep inelastic neutrino scattering events, as discussed in \secref{nubackgrounds}.

\subsection{Analysis Cuts for FASER$\nu$2}

The kinematic features of the signal suggest simple cuts on both $E_e$ and $\theta_e$ that can efficiently discriminate between the DM signal and the neutrino-induced backgrounds. In \tableref{cuts}, we show the cuts for FASER$\nu$2 employed in the analysis below.  These are sufficient to demonstrate interesting sensitivity reaches for our study, but we expect that even better background rejection could be achieved with a dedicated optimization procedure once the detector design is finalized.

\begin{table}
\begin{center}
\begin{tabular}{|c|c|}
\hline
Electron recoil energy & $300~\mev < E_e < 20~\gev$\\
\hline
\multirow{3}{*}{Electron recoil angle} & $10~\mrad < \theta_e < 20~\mrad$ for $E_e>10~\gev$\\
 & $10~\mrad < \theta_e < 30~\mrad$ for $3~\gev<E_e \le 10~\gev$\\
 & $10~\mrad < \theta_e$ for $E_e \le 3~\gev$\\
\hline
Track visibility & no additional charged tracks with $p>300~\mev$\\
\hline
\end{tabular}
\end{center}
\vspace*{-0.2in}
\caption{Analysis cuts used in the background analysis in \secsref{nubackgrounds}{mubackgrounds} for the emulsion detector FASER$\nu$2.\label{table:cuts}}
\end{table}

As we have discussed above, our sensitivity estimates below are obtained for a future emulsion detector similar to, but larger than, the FASER$\nu$ detector~\cite{Abreu:2019yak,Abreu:2020ddv} that will operate during LHC Run~3. This dictates the charged track visibility criteria in the emulsion. In our analysis, this will be modeled by a simple cut on the charged particle momentum, $p>300~\mev$. We employ this cut both when analyzing the DM signal events and the neutrino-induced backgrounds. In particular, we reject all background events characterized by additional visible charged tracks emerging from the vertex, other than a single electron or positron. When presenting the results in \secref{results}, we also discuss the impact of tightening the electron recoil energy cut on the expected sensitivity reach.

\subsection{Analysis Cuts for \flare} 

One of the main advantages of LArTPC detectors, which makes them excellent tools for probing interactions of medium-energy  neutrinos, is their capability to detect very soft charged particles produced in interactions in the detector. The relevant energy threshold can be as low as $25~\mev$ for the protons produced in neutrino scatterings off nuclei~\cite{MicroBooNE:2019knj}, while similar energy thresholds of $30~\mev$ can be achieved for muons and pions~\cite{Acciarri:2015uup}.  (See also Refs.~\cite{Ballett:2016opr,Batell:2019nwo} for recent discussions regarding new physics searches in the short-baseline neutrino detectors.) In the analysis below, we assume a $30~\mev$ threshold for the electron recoil energy~\cite{Acciarri:2015uup}. On the other hand, the background rejection capabilities are only mildly degraded by the poorer angular resolution of LArTPCs relative to emulsion detectors. We assume below that this resolution in LArTPCs is of the order of $1^\circ$ (see Ref.~\cite{Acciarri:2015uup}) and that we can cut on angles larger than $30~\mrad$. We summarize the cuts used in this analysis in \tableref{cutsLAr}. To understand the sensitivity of our results to the LArTPC angular resolution, in \secref{results}, we also present results assuming no cut on $\theta_e$, but more stringent cuts on the electron recoil energy, $30~\mev < E_e < 3~\gev$.

\begin{table}
\begin{center}
\begin{tabular}{|c|c|}
\hline
Electron recoil energy & $30~\mev < E_e < 20~\gev$\\
\hline
\multirow{2}{*}{Electron recoil angle} & $\theta_e < 30~\mrad$  for $E_e>3~\gev$\\
 & no $\theta_e$ cut for $E_e\leq 3~\gev$\\
\hline
Track visibility & no additional charged tracks with $p>30~\mev$\\
\hline
\end{tabular}
\end{center}
\vspace*{-0.2in}
\caption{Analysis cuts used in the background analysis in \secsref{nubackgrounds}{mubackgrounds} for the LArTPC detectors \flareten and \flarehundred.\label{table:cutsLAr}}
\end{table}

The rejection of both neutrino-induced and muon-induced backgrounds will further benefit from the capabilities of LArTPC detectors to disentangle single-electron tracks from $\gamma\to e^+e^-$ pair production. This can be done with high accuracy through the so-called ``$dE/dx$ discrimination method,'' i.e., by measuring the ionization at the beginning of the electromagnetic shower~\cite{Chen:2007ae}.  When discussing the muon-induced backgrounds in \secref{mubackgrounds}, we will assume that 85\% of these muon-induced $e^+e^-$ background events can be distinguished from single $e^-$ signal events~\cite{MicroBooNE:2019knj}.

\section{Neutrino-Induced Backgrounds}
\label{sec:nubackgrounds}

The signature of DM-electron scattering can most straightforwardly be mimicked by the interactions of SM neutrinos; see Ref.~\cite{Formaggio:2013kya} for a review.  These can either be due to $\nu$ scatterings off electrons characterized by the same event topology, or in much more abundant neutrino scatterings off nuclei that can occasionally resemble $\chi e^-\to\chi e^-$ signal events. We now estimate the expected event rates of these background processes and analyze how they can be reduced by employing simple kinematic cuts on the electron recoil energy and angle, as discussed in \secref{signal}. To this end, we simulate neutrino scattering events in the detector with the \texttt{GENIE} MC package~\cite{Andreopoulos:2009rq,Andreopoulos:2015wxa}.

\subsection{Neutrino Flux and Spectrum} 

The uncertainties in our background estimates are primarily related to the uncertainties in the modeling of the far-forward neutrino flux and energy spectrum at the LHC. In the future, the relevant simulations will greatly benefit from the neutrino data collected by FASER$\nu$ during LHC Run~3. In our analysis below we rely on the results obtained by the CERN Sources, Targets, and Interactions (STI) Group~\cite{Beni:2020yfy}, which utilized the FLUKA~\cite{Ferrari:2005zk,Battistoni:2015epi} model of the forward LHC optics and infrastructure. Notably, a similar FLUKA study of the muon flux going through the FASER location~\cite{FLUKAstudy} has been found to be in good agreement with initial measurements obtained during Run~2~\cite{Ariga:2018pin}. In our estimates, we take into account the full HL-LHC integrated luminosity of $\mathcal{L}=3~\iab$ and the FASER$\nu$2 and LArTPC detector designs discussed in \secref{signal}. We note that further analyses of the forward-neutrino flux and energy spectrum can be found, e.g., in Refs~\cite{Abreu:2019yak,Bai:2020ukz}.

\subsection{Neutrino-Electron Backgrounds}

\begin{figure}[t]
\begin{center}
\includegraphics*[width=1.0\textwidth]{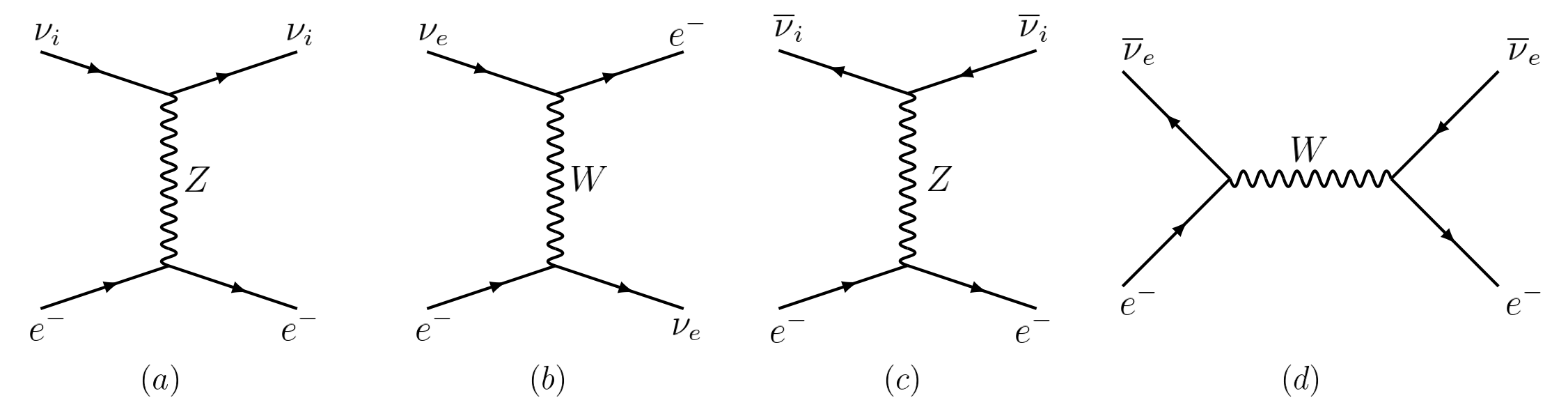}
\hfill
\end{center}
\vspace*{-0.3in}
\caption{Neutrino-electron elastic scattering background processes: $(a)$ NC $\nu_i e^- \to \nu_i e^-$, $(b)$ CC $\nu_e e^- \to \nu_e e^-$, $(c)$ NC $\overline\nu_i e^- \to \overline\nu_i e^-$, and $(d)$ CC $\overline\nu_e e^- \to \overline\nu_e e^-$.}
\label{fig:nu-e}
\end{figure}

The most important neutrino-induced backgrounds are neutrinos scattering with electrons in the detector, $\nu e^-\to \nu e^-$, as depicted in \figref{nu-e}. One expects $\mathcal{O}(10^2)$ such events in the FASER$\nu$2 and \flareten detectors during the entire HL-LHC phase and about $10^3$ events in \flarehundred. These events are topologically identical to the signal, and so must be reduced with kinematic cuts.

The most important such processes are associated with neutral current (NC) scatterings of muon neutrinos and similar interactions of electron neutrinos. Despite the lower flux of $\nu_e$s going through the detector, the corresponding scattering cross section has both NC and charged current (CC) contributions and is thus enhanced in comparison to the $\nu_\mu$ NC scattering cross section. On the other hand, tau neutrinos contribute negligibly to the expected backgrounds. We present the relevant estimates in \tableref{neutrinoelectronBGall} for all of the detector designs we consider.

\begin{table}
\begin{center}
\begin{tabular}{|c|c|} 
\hline
& $\nu e^-\to \nu e^-$ events after cuts\\
Detector  & (see Tables~\ref{table:cuts} and \ref{table:cutsLAr})\\
\hline
 FASER$\nu$2 (emulsion detector) & $10.1$ \\
 \flareten & $20.1$ \\
 \flarehundred & $94.7$ \\
 \flarehundred ($E_e<3~\gev$) & $15.7$ \\
 \hline
\end{tabular}
\end{center}
\vspace*{-0.2in}
\caption{The expected number of neutrino-electron scattering events during HL-LHC for the emulsion detector FASER$\nu$2 and for the \flareten and \flarehundred detectors, after applying the cuts shown in Tables~\ref{table:cuts} and \ref{table:cutsLAr}. For \flarehundred, we also show the expected number of background events assuming a more stringent cut on the electron recoil energy, $E_e<3~\gev$.
\label{table:neutrinoelectronBGall}}
\end{table}

In the case of FASER$\nu$2, we also show in \tableref{neutrinoelectronBG} a detailed breakdown of the number of events for different neutrino flavors after imposing successive cuts in the analysis. As can be seen, more than $90\%$ of the $\nu e^-\to\nu e^-$ background events can be rejected by applying the upper limit on the electron recoil energy, $E_e<20~\gev$. In contrast, the impact of such a low-recoil-energy cut on the DM signal events is typically much less severe, given the light mediator mass, as discussed in \secref{signal}. A small additional suppression of the background events comes from imposing the lower limit on the electron recoil angle, $\theta_e\agt 10~\mrad$, which slightly increases the signal to background ratio ($S/B$). The total number of expected such background events after cuts in FASER$\nu$2 is about $B_{\nu\text{-}e}\approx 10$. 

The last condition on the minimal recoil angle, however, is not crucial in the analysis. In particular, we do not impose such a cut when estimating the number of background events for the LArTPC detectors due to their limited angular resolution. As a result, the total rate of neutrino-electron scatterings in \flareten is at the level of $20$ expected background events in HL-HLC. This number grows to only about $100$ expected background events for \flarehundred. In this case, the larger total size of the detector is partly compensated by the relative increase of its transverse size and corresponding decrease of the neutrino flux per unit area away from the beam collision axis. 

\begin{table}
\begin{center}
\begin{tabular}{|c|c|c|c|} 
\hline
\boldmath{$\nu e^-\to \nu e^-$} &  & \multicolumn{2}{c|}{}\\
FASER$\mathbf{\nu}$2 & $E_e>300~\mev$ & \multicolumn{2}{c|}{$300~\mev < E_e < 20~\gev$}\\
(emulsion detector) &  & no cut on $\theta_e$ & $\theta_e$ cut (see \tableref{cuts})\\
\hline
 $\nu_e$ & $160$ & $4.7$ & $2.3$ \\ 
 $\bar{\nu}_e$ & $60$ & $4.3$ & $2.2$ \\ 
 $\nu_\mu$ & $70$ & $6.6$ & $3.2$ \\ 
 $\bar{\nu}_\mu$ & $44$ & $5$ & $2.4$ \\ 
 $\nu_\tau$ & $1$ & $0.05$ & $0.02$ \\ 
 $\bar{\nu}_\tau$ & $1$ & $0.04$ & $0.02$ \\ 
 \hline
 \textbf{TOTAL} & $336$ & $20.7$ & $\mathbf{10.1}$ \\
 \hline
\end{tabular}
\end{center}
\vspace*{-0.2in}
\caption{The expected number of neutrino-electron scattering events during HL-LHC for the emulsion detector FASER$\nu$2 for each neutrino and anti-neutrino flavor. Successive columns correspond to the expected number of such events after imposing the cuts discussed in the text. 
\label{table:neutrinoelectronBG}}
\end{table}

\subsection{Neutrino-Nuclei Backgrounds}

\begin{figure}[t]
\begin{center}
\includegraphics*[width=1.0\textwidth]{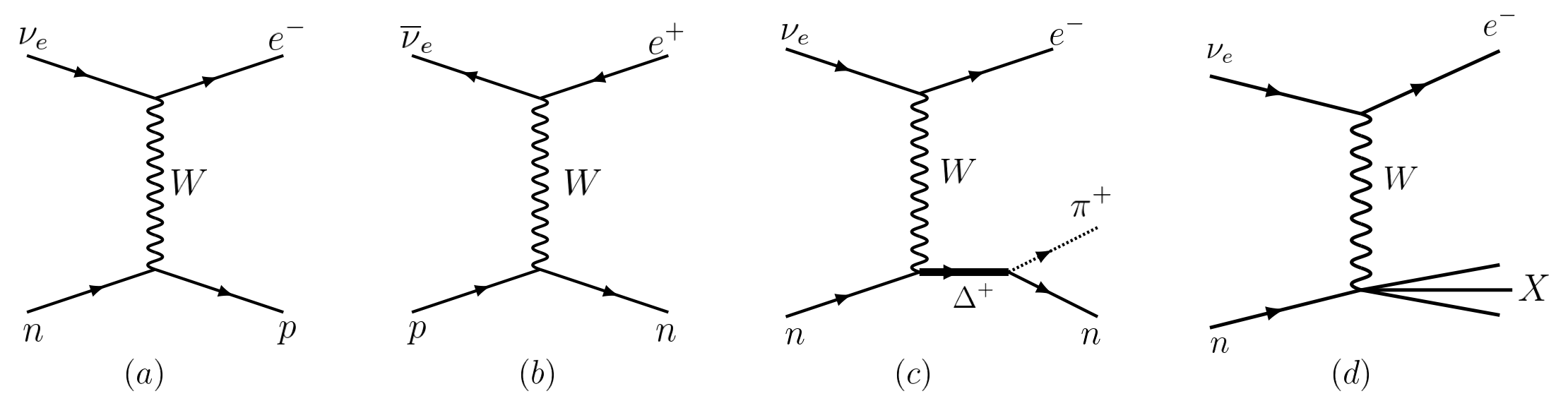}
\hfill
\end{center}
\vspace*{-0.3in}
\caption{Neutrino-nuclei background processes: $(a)$ neutrino CCQE scattering $\nu_e n \to e^- p$, 
$(b)$ anti-neutrino CCQE scattering $\overline\nu_e p \to e^+ n$, and representative $(c)$ CCRES and $(d)$ CCDIS reactions. }
\label{fig:nu-N}
\end{figure}

The total number of neutrino-nuclei scattering events significantly exceeds that of neutrino scatterings off electrons. Even when the electron recoil energy range is limited, as is dictated by the cuts used in our analysis, a few thousand of such events are still expected during the HL-LHC phase for the $10$-tonne detectors under study. These are mostly due to electron neutrino CC deep inelastic scattering (CCDIS), but there are also important contributions from CC quasi-elastic (CCQE) and resonant (CCRES) interactions; see \figref{nu-N}. We now discuss these backgrounds, first for the emulsion detector FASER$\nu$2, and then for the LArTPC detectors \flare.

The CCQE scattering processes, $\nu_e n \to e^- p$ and $\bar{\nu}_e p \to e^+ n$, are characterized by a high-energy electron in the final state and soft activity from the nuclear recoil. Hence, if the outgoing electron satisfies the cuts, such events can easily resemble DM-electron scatterings. This is especially true for events with a neutron in the final state, as it does not leave a track in the emulsion and can travel a distance of the order of the hadronic interaction length in tungsten, $\lambda_{\textrm{had,W}} \approx 10~\cm$~\cite{Zyla:2020zbs}, before interacting. Events with a final-state proton can also mimic $\chi e^- \to\chi e^-$ scatterings if the emitted proton is too soft or if it is reabsorbed before leaving the nucleus due to final state interactions (FSI).

In CCQE scattering events, most of the neutrino energy is transferred to the outgoing electron. Therefore, this background contribution is associated with lower energy neutrinos, $E_\nu \alt 20~\gev$, given our cut on the electron recoil energy. As shown in \tableref{CCQE}, we expect about $10$ such background events during the entire run of FASER$\nu$2. However, these events are typically associated with a larger electron recoil angle $\theta_e$ and can be effectively rejected by selecting events with low recoil angle. Last, but not least, the cut on the additional energetic charged tracks emerging from the vertex further suppresses this background. This suppression is more pronounced for $\nu_e$ scattering events with protons in the final state. Both the neutrino and anti-neutrino CCQE scatterings lead to less than one expected background event.

\begin{table}[t]
\begin{center}
\begin{tabular}{ |c|c|c|c| }
\hline
\textbf{CCQE} & \multicolumn{3}{c|}{$300~\mev < E_e < 20~\gev$}\\
FASER$\nu$2 & no cut on $\theta_e$ & \multicolumn{2}{l|}{$\theta_e$ cut (see \tableref{cuts})} \\
(emulsion detector) & & & no other charged tracks with $p>300~\mev$\\
 \hline
 $\nu_e$ & $9$ & $0.45$ & $0.2$\\ 
 $\bar{\nu}_e$ & $5$ & $0.25$ & $0.2$\\ 
 \hline
 \textbf{Total} & $14$ & $0.7$ & $\mathbf{0.4}$\\
 \hline
\end{tabular}
\end{center}
\vspace*{-0.2in}
\caption{The expected number of CCQE background events during HL-LHC for the emulsion detector FASER$\nu$2.\label{table:CCQE}}
\end{table}

\begin{table}[!h]
\begin{center}
\begin{tabular}{ |c|c|c|c| }
\hline
\textbf{CCRES} & \multicolumn{3}{c|}{$300~\mev < E_e < 20~\gev$}\\
FASER$\nu$2 & no cut on $\theta_e$ & \multicolumn{2}{l|}{$\theta_e$ cut (see \tableref{cuts})} \\
(emulsion detector) & & & no other charged tracks with $p>300~\mev$\\
 \hline
 $\nu_e$ & $22$ & $0.9$ & $0.2$\\ 
 $\bar{\nu}_e$ & $14$ & $0.8$ & $0.5$\\ 
 \hline
 \textbf{Total} & $36$ & $1.7$ & $\mathbf{0.7}$\\
 \hline
\end{tabular}
\end{center}
\vspace*{-0.2in}
\caption{The expected number of CCRES background events during HL-LHC for the emulsion detector FASER$\nu$2.\label{table:CCRES}}
\end{table}

\begin{table}[!h]
\begin{center}
\begin{tabular}{ |c|c|c|c| }
\hline
\textbf{CCDIS} & \multicolumn{3}{c|}{$300~\mev < E_e < 20~\gev$}\\
FASER$\nu$2 & no cut on $\theta_e$ &  \multicolumn{2}{l|}{$\theta_e$ cut (see \tableref{cuts})}\\
(emulsion detector) & & & no other charged tracks with $p>300~\mev$\\
 \hline
 $\nu_e$ & $5.3$k & $40$ & $0.03$\\ 
 $\bar{\nu}_e$ & $1.2$k & $30$ & $0.06$\\ 
 \hline
 \textbf{Total} & $6.5$k & $70$ & $\mathbf{0.1}$\\
 \hline
\end{tabular}
\end{center}
\vspace*{-0.2in}
\caption{The expected number of CCDIS background events during HL-LHC for FASER$\nu$2.\label{table:CCDIS}}
\end{table}

Another source of background is resonant pion production in neutrino scatterings off nucleons (CCRES). After summing over all intermediate nuclear resonances, the inclusive CCRES scattering cross section can exceed that of the CCQE scatterings in the energy range relevant for our analysis. Fortunately, most of these events can again be disentangled from the DM signal based on the electron recoil angle. In addition, if the final state charged pions escape the nucleus, such events can be efficiently rejected by observing their charged tracks in the emulsion. This results in $B_{\textrm{CCRES}}\approx 0.7$ such background events, as shown in \tableref{CCRES}.

In the case of CCDIS events, even high energy neutrinos with $E_\nu\sim 100~\gev$ can produce outgoing electrons with $E_e<20~\gev$. On the other hand, the large momentum transfer to nuclei in CCDIS scatterings generally leads to additional energetic charged tracks emerging from the vertex. As a result, only neutrinos with energies that do not greatly exceed the $20~\gev$ threshold can effectively mimic the DM signal events. A combination of cuts on the additional visible charged tracks and on the electron recoil angle reduces the CCDIS backgrounds to $B_{\textrm{CCDIS}}\alt 0.1$ expected event, as shown in \tableref{CCDIS}.

The cuts that we use for the LArTPC detectors (see \secref{signal}) offer even better background rejection capabilities for neutrino-nuclei scattering events. The angular resolution of order $30~\mrad$ is sufficient to achieve a similar rejection power as in the case of the emulsion detector. At the same time, the lower energy threshold for charged particle detection implies much better identification of background events with additional tracks emerging from the interaction vertex.  Hence, we expect background from neutrino-nuclei scatterings to contribute with less than a single event for \flareten, and up to a handful of events for \flarehundred.

We briefly comment on two other classes of neutrino scattering reactions. First, the CC interactions of muon or tau neutrinos will typically be disentangled from the $\chi e^- \to \chi e^-$ signal by identifying the outgoing lepton. Second, neutrino-nuclei NC scattering may mimic the signal if a photon is emitted from the interaction vertex and is subsequently reconstructed in the emulsion as a single electron that satisfies all the cuts. In addition, all other visible charged tracks associated with the nuclear interaction vertex must be sufficiently soft to escape identification. The low likelihood of these combined circumstances suggests that these backgrounds are subdominant relative to the $\nu_e$-nucleus CC backgrounds discussed above, and so not of great concern.  We note, however, that photons imitating the single-electron signature in emulsion will be discussed further in \secref{mubackgrounds} in the context of muon-induced backgrounds.

\subsection{Neutrino-Induced Background Summary} 

Summing over all the neutrino-induced backgrounds, we expect $\mathcal{O}(10)$ such background events during the entire HL-LHC run for the FASER$\nu$2 detector, given the kinematic cuts discussed in \secref{signal}. For \flareten, the expected background is about a factor of two larger, given the less stringent cut on low values of $\theta_e$, and the number of background events for \flarehundred is about 100.  These backgrounds are dominated by neutrino-electron scattering, while neutrino-nuclei scatterings, although larger {\em a priori}, can be more efficiently disentangled from the signal in the analysis. 

\begin{figure}[tb]
\begin{center}
\includegraphics*[width=0.48\textwidth]{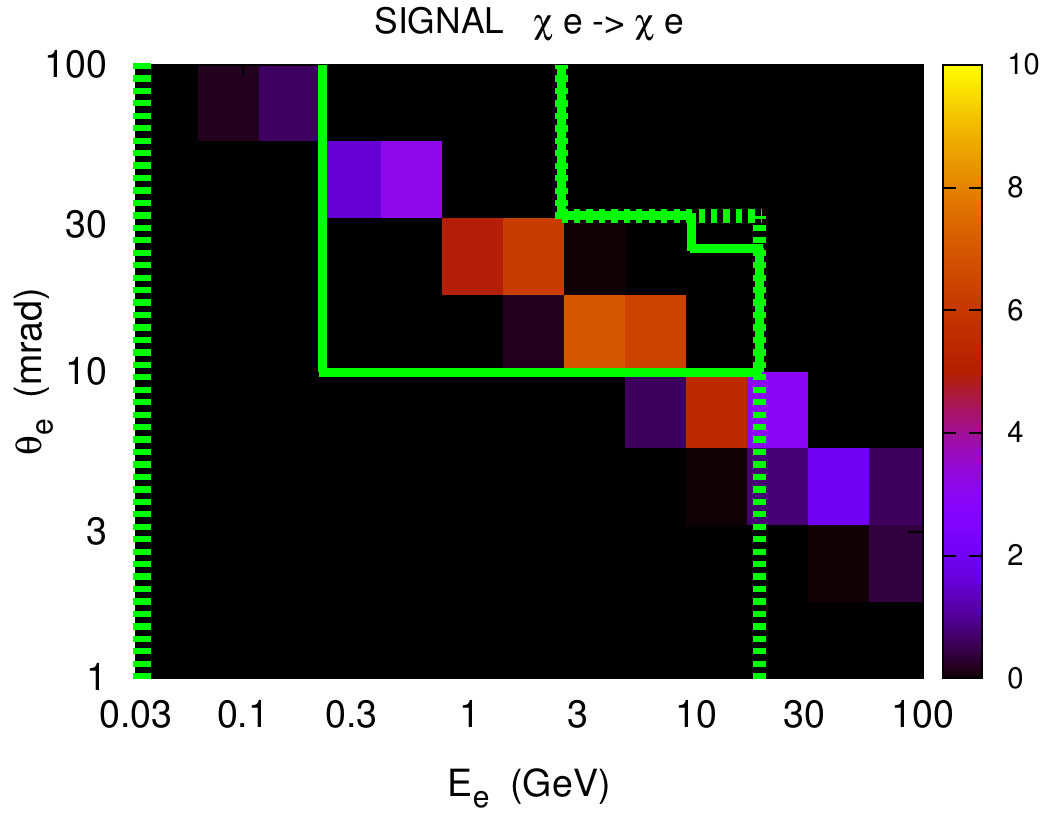} \\
\includegraphics*[width=0.48\textwidth]{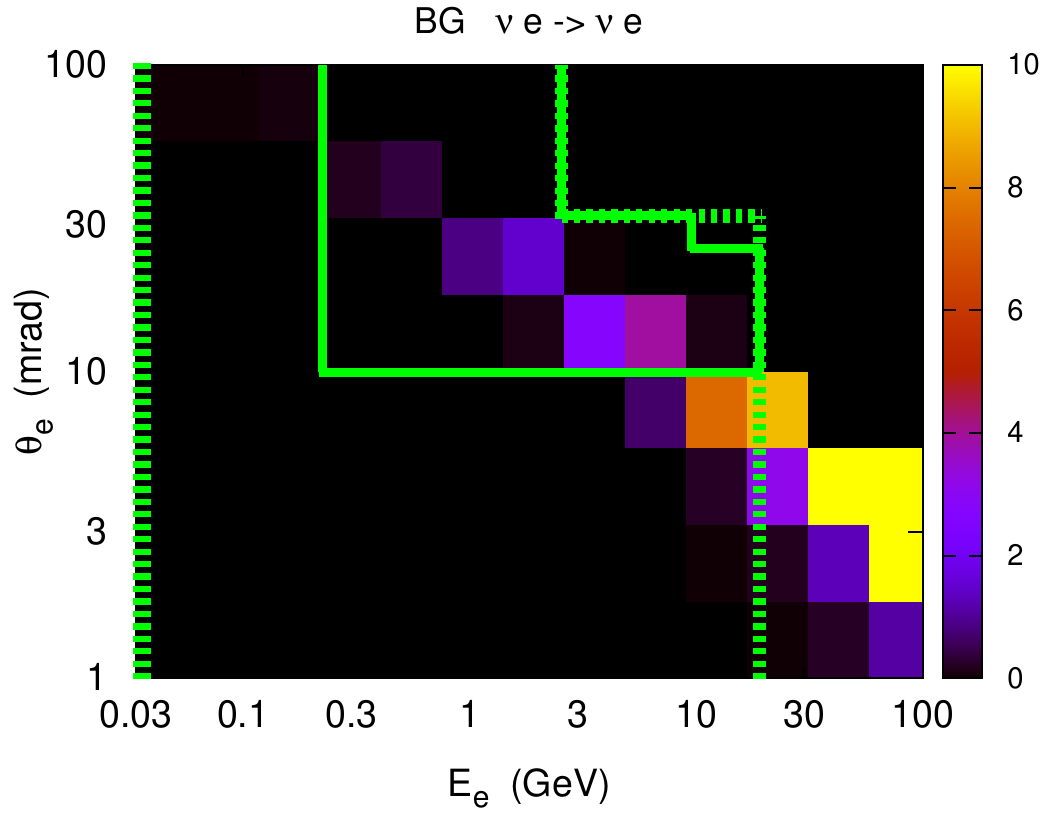}
\hfill
\includegraphics*[width=0.48\textwidth]{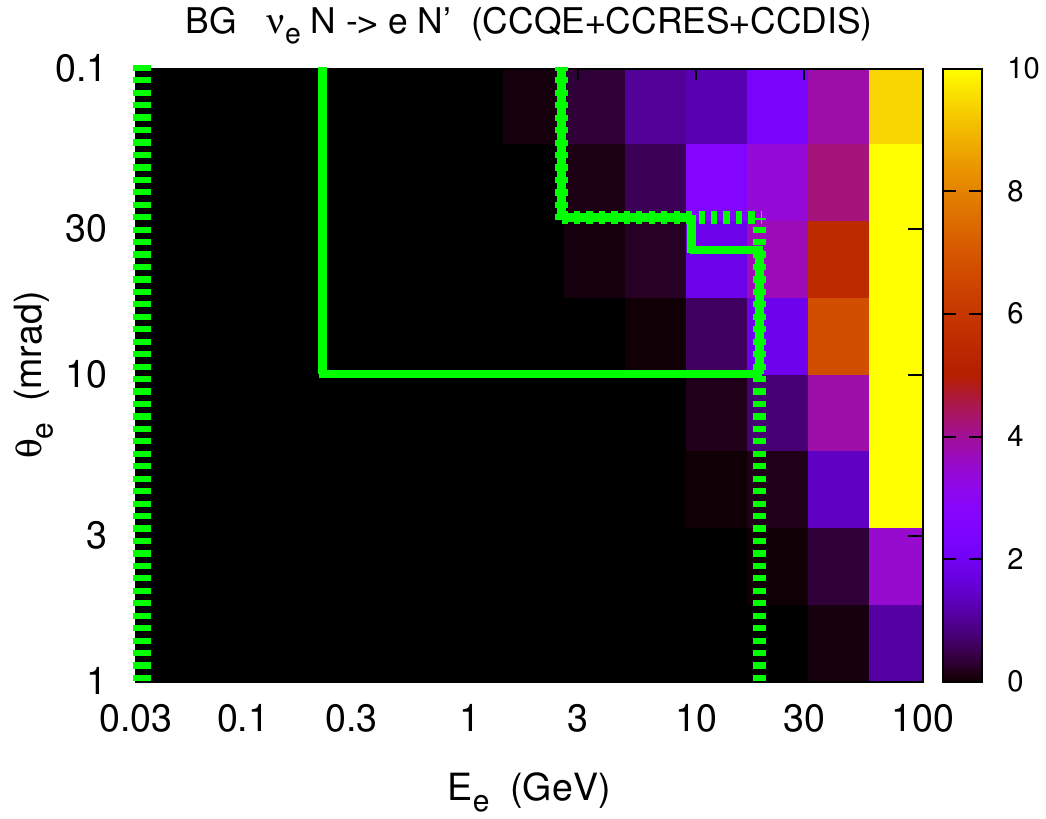}
\end{center}
\vspace*{-0.3in}
\caption{Event distributions in the $(E_e, \theta_e)$ plane, along with the regions selected by the cuts, for the emulsion detector FASER$\nu$2. We show these distributions for the DM signal, $\chi e^- \to\chi e^-$, for the benchmark scenario with $m_{A^\prime}= 3m_\chi = 75~\mev$, $\epsilon = 10^{-4}$ and $\alpha_D=0.5$ (top), and for the background from neutrino-electron (bottom left) and neutrino-nuclei (bottom right) scattering. The last distribution contains contributions from CCQE, CCRES and CCDIS scatterings. Only events with no additional charged tracks with $p>300~\mev$ emerging from the vertex, besides a single electron or positron, are shown. The solid and dashed green lines outline the signal region for the FASER$\nu$2 and LArTPC detectors \flare, respectively.  Note that for the LArTPC detectors, the number of signal and background events in the colorful bins will be different from those shown. 
}
\label{fig:heatmaps}
\end{figure}

The impact of the analysis cuts is illustrated in the $(E_e, \theta_e)$ plane in  \figref{heatmaps} for FASER$\nu$2. As discussed in \secref{model}, the DM interactions typically produce low-energy electron recoils, while $\nu$-$e$ interactions generate $E_e$ that is more spread over the entire available energy range. The $\nu$-$N$ interactions are instead characterized by electrons with both larger energy and larger recoil angle. Importantly, in the plots, we show only the background events for which there is only a single $e^\pm$ charged track emerging from the vertex with $p>300~\mev$. This greatly reduces the number of CCDIS background events. In \figref{heatmaps} we also show the signal regions for both the emulsion and LArTPC detectors that were defined in \secref{signal}, in which the excess of DM scattering events over the expected neutrino-induced background can most easily be seen.

\section{Muon-Induced Backgrounds\label{sec:mubackgrounds}}

\begin{figure}[t]
\begin{center}
\includegraphics*[width=0.55\textwidth]{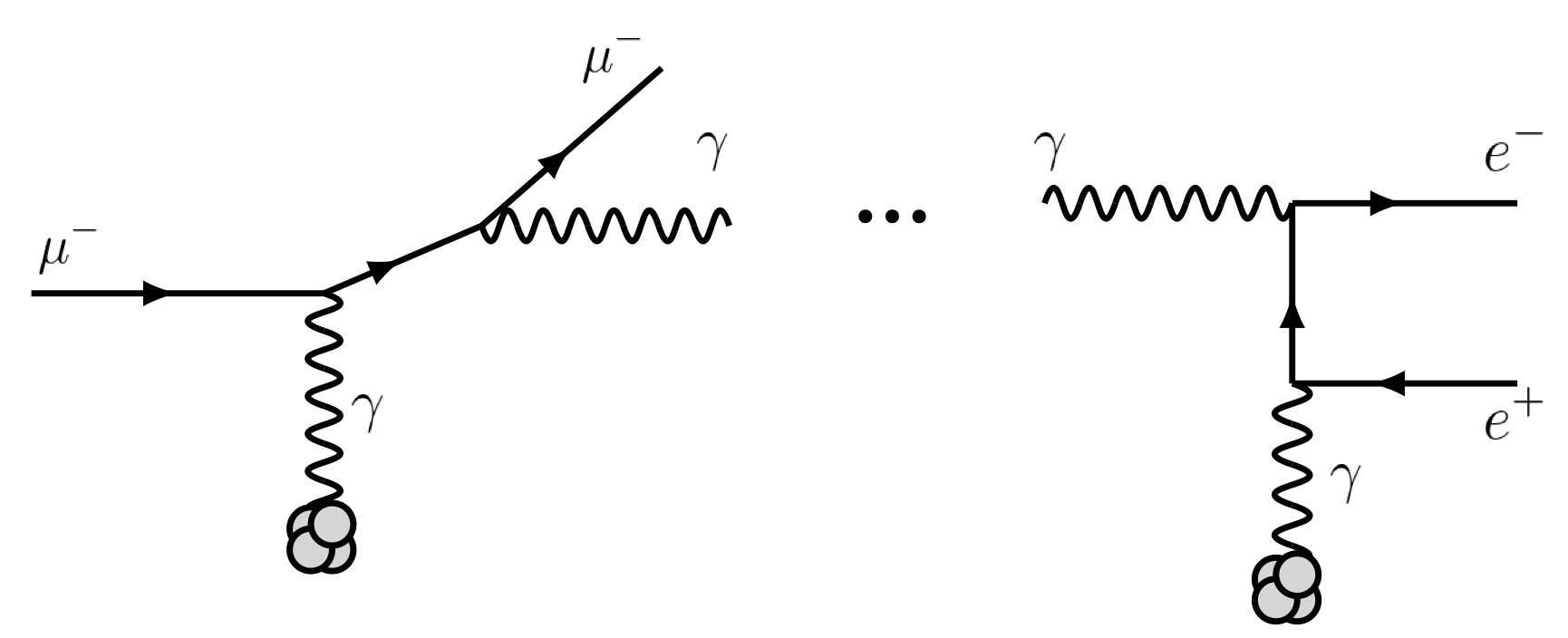}
\hfill
\end{center}
\vspace*{-0.3in}
\caption{Muon-induced background processes. A photon is produced via muon-bremsstrahlung and then converts to an $e^+e^-$ pair. Such an event is a background to the DM signal if the incoming muon is not associated with the $e^+e^-$ pair production, either the electron or positron has an energy below the detection threshold energy, and the other has the correct kinematics to pass the selection cuts. In addition to muon-bremsstrahlung, such backgrounds can arise in direct electron-positron pair production, $\mu N\to \mu e^+e^- N$.
\label{fig:muon-BG}}
\end{figure}

Aside from neutrinos, the only other SM particles that can pass through 100 m of rock are muons. These muons are dominantly produced either at the IP or through collisions in the TAXN neutral particle absorber. The relevant flux of muons going through a small FASER$\nu$ detector during LHC Run~3 has been predicted by the CERN STI group~\cite{FLUKAstudy} employing the FLUKA transport code~\cite{Ferrari:2005zk,Battistoni:2015epi} and was independently measured by the FASER collaboration during Run~2~\cite{Ariga:2018pin}. When rescaled to account for the HL-LHC luminosity and the larger transverse size of FASER$\nu$2, more than $10^{11}$ muons are expected to traverse the detector during the entire run of the experiment, and this number grows by a factor of a few for the LArTPC detectors. If not removed through selection cuts or through a dedicated muon veto, muon-induced photons that convert to electron-positron pairs inside the detector can occasionally mimic the DM signal, as illustrated in \figref{muon-BG}. Below, we briefly discuss such backgrounds and possible strategies to mitigate their impact on the DM search.

\subsection{Sweeper Magnet} 

Part of the muon-induced background can be removed by installing a sweeper magnet placed along the beam collision axis upstream of the detector. A convenient location for such a magnet is along the beam collision axis after it has left the LHC beampipe, but before it enters the tunnel wall. This location is roughly 300 m from the ATLAS IP and 200 m upstream of the detector.  A magnet placed in this location will not interfere with transport, which typically uses the wider path on the other side of the LHC beampipe.  

A muon with energy $E_{\mu}$ that travels a distance $\ell$ through a magnetic field $B$ oriented perpendicular to its direction and then travels an additional distance $d$ is deflected by a distance 
\begin{equation}
h_{B} \approx \frac{ec d  }{E_{\mu}} B \ell = 60~\cm  
\left[\frac{100~\gev}{E_{\mu}} \right] \left[\frac{d}{200~\m} \right] 
\left[\frac{B \cdot \ell }{\text{T} \cdot \m } \right] 
\end{equation}
in the transverse plane.  Permanent dipole magnets with a magnetic field of $B \approx 0.57~\text{T}$ and total length of 3.5 m for an integrated magnetic field strength of $B \cdot \ell = 2.0~\text{T} \cdot \m$ have already been constructed for the FASER experiment.  These magnets require no services and have very small fringe fields.  It is also important that the sweeper magnet accommodates shifts in the beam collision axis from variations in the beam crossing angle.  At the HL-LHC, the beam crossing angle may vary by up to 590 $\mu$rad, corresponding to a shift of 9 cm at a distance of 300 m from the IP.  The FASER magnets have an inner diameter of 20 cm, which is roughly of the size required to accommodate such shifts.  

It therefore appears quite feasible for a sweeper magnet to eliminate all muons with energies below 100 GeV from the detector region. In fact, with a ten-fold increase in $B\cdot \ell$ of the sweeper magnet with respect to the current FASER magnets, one could even deflect muons with few $\tev$ energies, drastically reducing the number of muon-induced backgrounds.  Of course, in a detailed study, it would be important to trace all muons through the beam optics and be sure that, in sweeping away muons from the detector one does not simultaneously sweep other muons into the detector. 

\subsection{Single-Electron-Like Events from Muon Interactions} 

As mentioned above, even if a sweeper magnet is used, some of the muons could be deflected back into the detector. Notably, the path of such deflected muons may often not be parallel to the beam collision axis. Given the angular cuts used in the DM search, as described in \secref{signal}, this could greatly suppress such muon-induced backgrounds. On the other hand, if the sweeper magnet deflects only a portion of the high-energy muons, the remaining ones will traverse  the detector with only minimal deflection angles that, in the first approximation, can be neglected in the analysis. It is then useful to analyze the potential impact of such muons on the DM search. In the rest of this section, we conservatively assume that only muons with energies smaller than $100~\gev$ will be swept away, while the more energetic ones will pass through the detector.

\begin{figure}[t]
\begin{center}
\includegraphics*[width=0.45\textwidth]{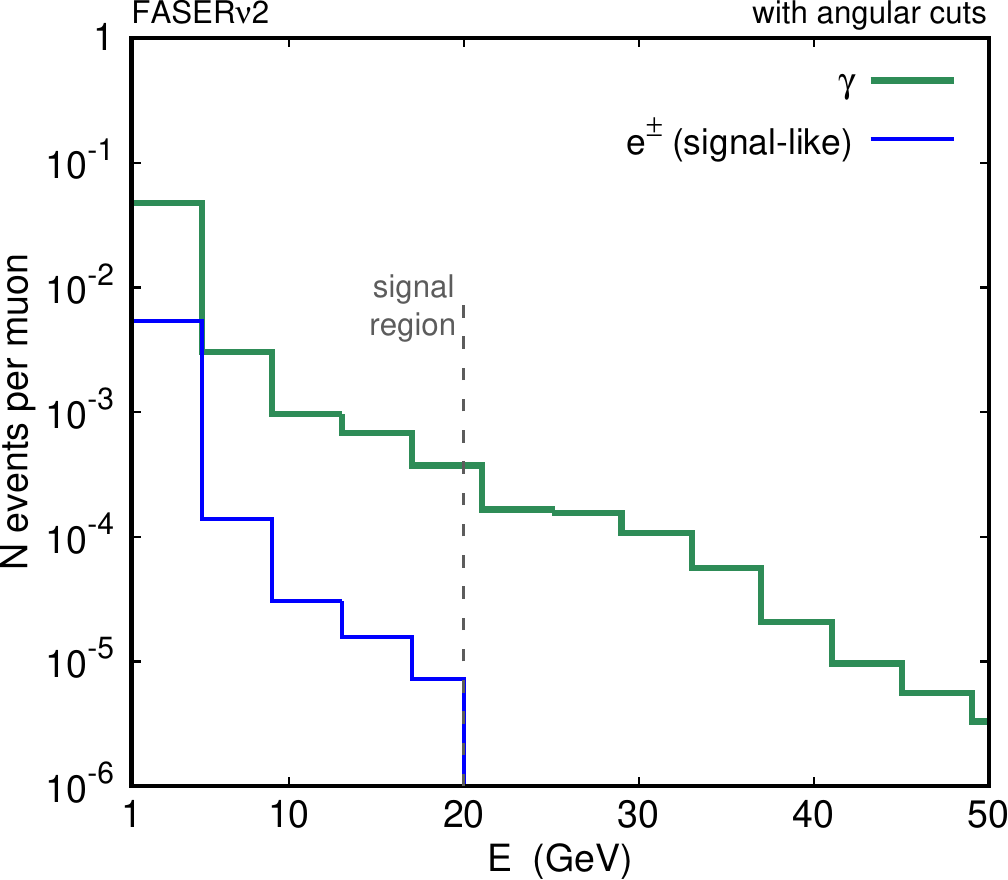}
\end{center}
\vspace*{-0.3in}
\caption{The expected energy spectrum of muon-induced photons (green) in FASER$\nu$2 during HL-LHC that emerge from the $1$ mm thick tungsten layers. In the plot, the angle $\theta_{\gamma}$ with respect to the parent muon direction satisfies the cuts introduced in \secref{signal}. For the estimates, the parent muon spectrum with $E_\mu>100~\gev$ has been used, following Refs.~\cite{FLUKAstudy,Abreu:2019yak}. The spectrum of single electrons or positrons that can mimic the DM signal after photon pair production is also shown (blue). These are events in which a muon-induced photon pair produces, and either the electron has an energy below the detectability threshold and the positron satisfies the kinematic cuts discussed in Table~\ref{table:cuts}, or vice versa. The signal region in the DM search corresponds to $300~\mev<E_e<20~\gev$ as indicated in the plot.}
\label{fig:muonBG}
\end{figure}

For such high-energy muons, the photon bremsstrahlung and $e^+e^-$ pair production interaction lengths are of the order of a few meters for tungsten and above $100~\m$ for liquid argon~\cite{Groom:2001kq}. This produces a large number of photons and $e^+e^-$ pairs in both types of detectors. For illustration, in \figref{muonBG} we show the expected energy spectra for FASER$\nu$2. To obtain these spectra, we performed dedicated FLUKA simulations for this study, starting with the parent muon high-energy ($E_\mu>100~\gev$) spectrum predicted for LHC Run~3, and requiring that the resulting photon or $e^+ e^-$ pair have angles with respect to the beam axis satisfying the cuts discussed in \secref{signal}. The muon-induced photons will typically produce an $e^+ e^-$ pair within the radiation length of the parent muon, which is equal to $\lambda_{\textrm{W}} = 0.35~\cm$ and $\lambda_{\textrm{LAr}} = 14~\cm$ in tungsten and liquid argon, respectively~\cite{Zyla:2020zbs}.

We first focus on the emulsion detector. Given the high density of tracks in the emulsion, this distance is often too large to associate the secondary photon pair-production vertex, $\gamma N\to e^+e^- N$, with the parent muon. If sufficiently energetic, the $e^-$ and $e^+$ tracks may always be differentiated as two separate tracks in an emulsion detector, given its extraordinary resolution. However, these events are a background to the $\chi e^- \to \chi e^- $ signal when either the electron or positron has an energy below the 300 MeV threshold energy and thus evades detection, while the other has the energy and direction required to pass the cuts. The predicted spectrum of such events is shown in the green histogram of \figref{muonBG}. As can be seen, up to 1\% of muons going through the detector can lead to such events, although this fraction drops to about $0.01\%$ for increased energy threshold $E_e>3-5~\gev$. 

Other processes may also lead to muon-induced background events. In particular, even if the photon pair produces the electron and positron with both the energies above the visibility threshold, one of them can immediately radiate a photon in the same tungsten layer, $e N\to \gamma e N$. The outgoing electron/positron might then emerge from this layer as either too soft or significantly deflected, which affects the photon vertex identification. Similar processes can lead to background events after direct $e^+e^-$ pair production, $\mu N\to \mu e^+e^- N$, which dominates the soft muon-induced $e^\pm$ spectrum. In this case, however, the visible signal-like $e^\pm$ track is produced in the close vicinity of the parent muon track, which allows for better background rejection.

The corresponding expected number of bremsstrahlung processes per muon in the liquid-argon \flareten detector is about an order of magnitude smaller than for FASER$\nu$2. This is due to the lower density and the smaller value of the radiative muon energy-loss function $b(E)$ in this material~\cite{Groom:2001kq}. On the other hand, this is partly offset by both the lower detection energy threshold and the absence of a lower cut on the emission angle, resulting in a substantial number of low-energy bremsstrahlung photons that could mimic the DM signature. We expect that most of these photons can be disentangled from the single-electron-initiated EM showers using the $dE/dx$ discrimination method mentioned in \secref{signal}, while still $\mathcal{O}(10\%)$ could be misreconstructed as signal-like events. As a result, one finds less than $1\%$ of muons are able to generate signal-like events in \flareten. A similar conclusion holds for \flarehundred.

Notably, one could also consider a modified design of \flarehundred with a larger transverse size and reduced length in comparison with \eqref{LAr100ton}. In this case, an additional important effect could come from the non-uniform distribution of muons in the transverse plane at the location of the experiment. In particular, for off-axis locations with radial distances of about $R\sim 2~\m$ from the beam collision axis, the flux of through-going muons can grow by up to two orders of magnitude depending on the direction in the transverse plane; see Refs.~\cite{Ariga:2018pin,FLUKAstudy}. This is due to the bending of muons by the LHC magnets and, in precise modeling, would have to be considered along with the possible impact of the aforementioned sweeping magnet. If not deflected with the help of a properly designed sweeping magnet, the larger muon flux in outer parts of the detector could necessitate an effective reduction of the fiducial volume used in the DM search.

\subsection{Active Muon Veto for FASER$\nu$2} 

In the absence of perfect muon sweeping, another strategy is to actively veto muon-induced background events. This could be achieved by using timing information about both the through-going muon and the EM shower in the emulsion. 

The timing information about the EM showers in the emulsion could be obtained by interleaving the emulsion detector with electronic tracker layers, similar to the proposed designs of SND@SHiP~\cite{Anelli:2015pba} and SND@LHC~\cite{Ahdida:2020evc}, as well as that employed in the OPERA emulsion detector~\cite{Acquafredda:2009zz}. A distance between the layers of order $10-15$ radiation lengths will allow one to successfully observe most of the EM showers that will typically leak to the electronic detectors. A too-large separation between the layers would mean a reduced effective volume of the detector, as only a fraction of the showers will be registered, limiting the prospects of the DM search. 

Once the candidate DM signal event in the electronic detector is identified with no time-coincident muon, further analysis would be based on proper matching between the activity in the tracker and the EM vertex in the emulsion. Such matching could suffer from the pile-up of numerous partially overlapping EM showers in the emulsion. This will have to be overcome when scanning the emulsion in search of candidate events. The detailed analysis of this issue and the development of the algorithms used to identify the EM showers in the emulsion is left for future studies, but see Ref.~\cite{Juget:2009zz} for a discussion about EM shower reconstruction in the OPERA experiment.

\subsection{Active Muon Veto for \flare} 

The muon-induced backgrounds can be more easily rejected in the LArTPC detectors. This is primarily due their capability to provide active time information about the events, which could be significantly enhanced by the use of the additional light collection system, as discussed in \secref{Detectors}. Combined with the decent spatial resolution offered by TPCs, this allows for efficient rejection of background events that can be associated with a time-coincident muon passing through the front veto and detector.

To identify the DM signal, one would first employ the excellent $\mathcal{O}(10~\ns)$ time resolution of the light collection system. Based on this, one can single out individual $\mathcal{O}(\textrm{ms})$ time windows in the TPC data that contain candidate events. Such events can then be differentiated from all neutrino-induced backgrounds based on their detailed characteristics, as measured by the TPC and the analysis cuts discussed in \secref{signal}. Importantly, given the low number of expected DM and neutrino scatterings, there is a negligible probability that both types of events to happen in a single time window. 

The muon-induced background is much more common. In the case of \flareten, we expect about $10$ muons passing the detector in each TPC time window. This is determined by the transverse size of the detector and the estimated muon flux of $1~\textrm{Hz}/\textrm{cm}^2$~\cite{FLUKAstudy}. Since muon-induced photons could mimic the single-electron-initiated EM showers, they would have to be rejected based on the spatial information about the event. In particular, given the aforementioned large radiation length in liquid argon and the maximum angle of the photon with respect to the beam-collision axis to mimic the signal, $\theta_{\textrm{max}} = 30~\mrad$, reducing the fiducial volume of the detector by several $\cm$ wide cylinders around each of the muon trajectories will be sufficient to effectively reject all such muon-induced backgrounds. At the same time, the DM signal detection rate will only be mildly affected by the order $10\%$ fiducial volume reduction, which has an almost imperceptible impact on the sensitivity reach plots presented in \secref{results}.

The muon-induced backgrounds could similarly be rejected in the larger \flarehundred detector. Note, however, that if the design of \flarehundred were changed with respect to \eqref{LAr100ton} by increasing its transverse size and reducing its length, the outer parts of the detector could be effectively eliminated from the analysis due to the increased muon flux, as discussed above. This would have to be taken into account when designing the optimized detector geometry.

\section{Results}
\label{sec:results}

In this section we present the sensitivity of these experiments to the dark photon-mediated DM models introduced in \secref{model} during the HL-LHC phase. In deriving our sensitivity projections, we will assume that muon-induced backgrounds can be reduced to a negligible level by making use of event time information, as outlined in \secref{mubackgrounds}. We will, however, take into account the irreducible neutrino-induced backgrounds discussed in \secref{nubackgrounds}.  It is conceivable that by measuring neutrino scattering processes in kinematic regions outside the DM signal region, precise measurements of the neutrino fluxes and improved modeling of neutrino scattering cross sections could be obtained during the eventual operation of these experiments. Such a data-driven approach may help to mitigate systematic uncertainties in the neutrino-induced background rates for the DM search. We will thus work under the assumption that statistical uncertainties in the single electron data sample dominate over systematic uncertainties. We begin the discussion with our results for the emulsion detector FASER$\nu$2, and then discuss our results for the LArTPCs \flareten and \flarehundred.

\subsection{Results for FASER$\nu$2} 

\begin{figure}[t]
\begin{center}
\includegraphics*[width=0.49\textwidth]{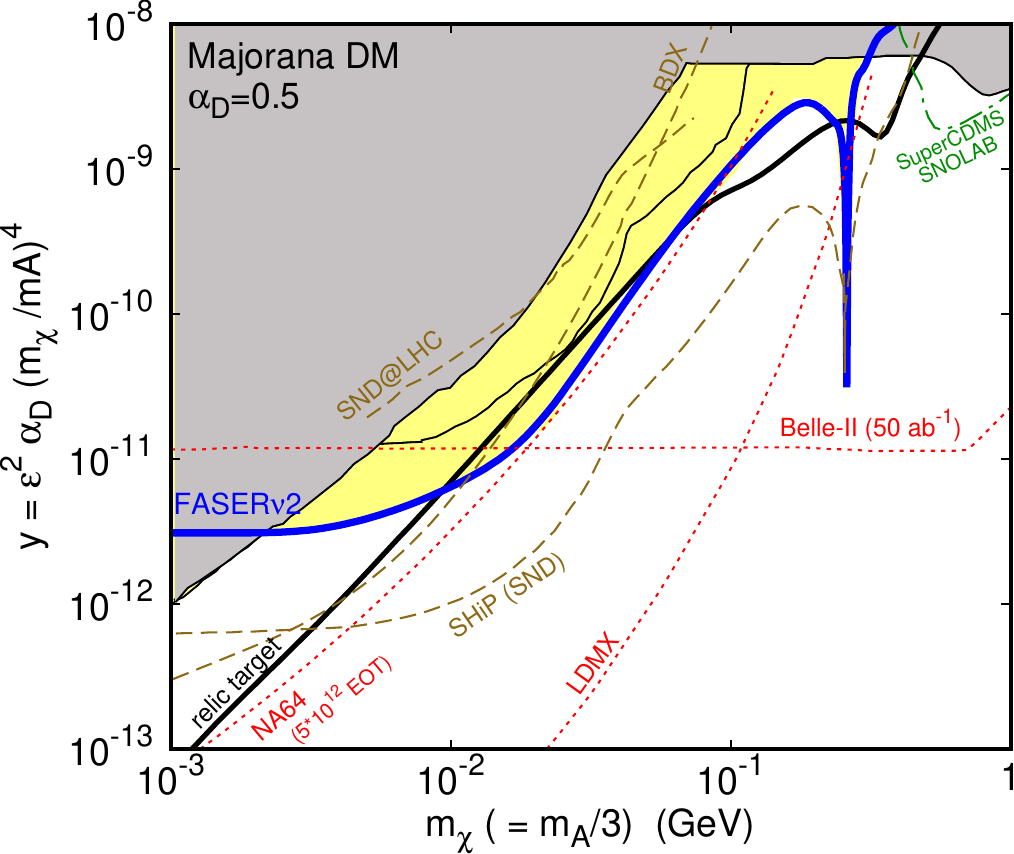}
\hfill
\includegraphics*[width=0.49\textwidth]{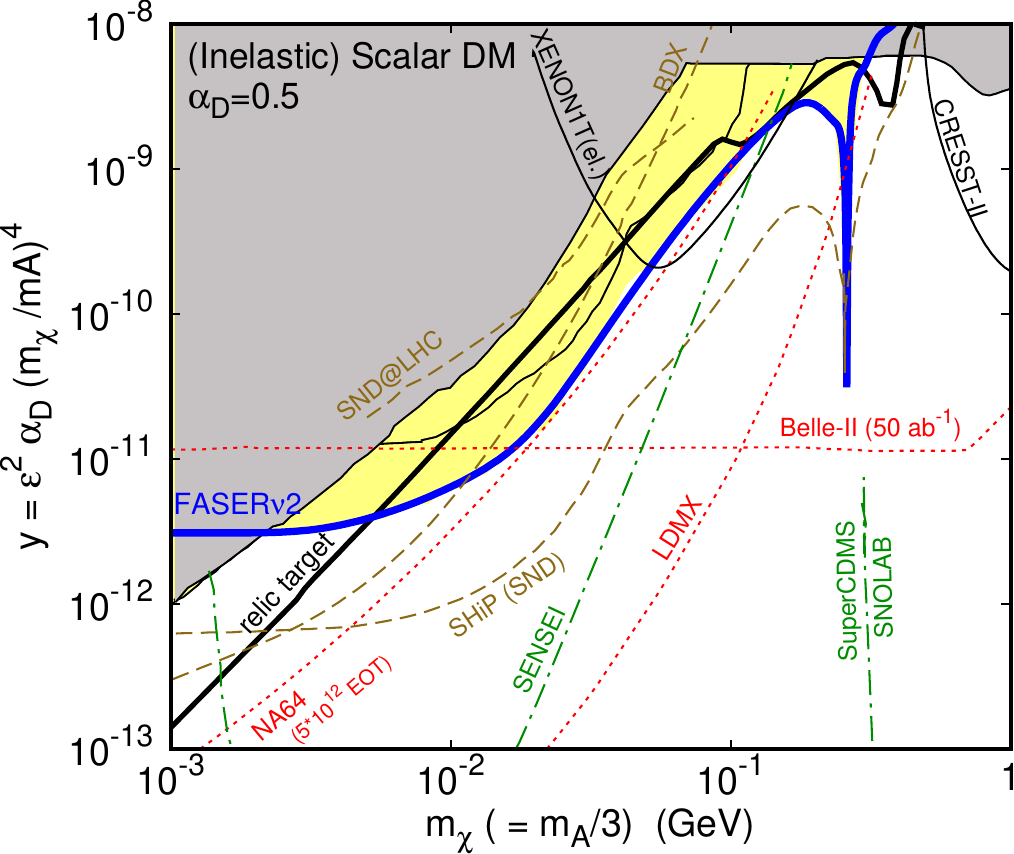}
\end{center}
\vspace*{-0.3in}
\caption{Projected $90\%$ CL exclusion bounds in the DM-electron scattering search at FASER$\nu$2 are shown with the blue lines and a yellow-shaded regions. The results are shown for $\alpha_D=0.5$, $m_{A^\prime}/m_\chi = 3$, and Majorana DM (left) and (inelastic) scalar DM (right). The solid black lines are the relic targets, where the DM has the correct thermal relic density. Current bounds are shown with gray-shaded regions and thin solid black lines (see the text for details). We also show with red dotted lines the projected sensitivities of future missing energy and momentum searches at Belle-II~($50~\ifb$)~\cite{Battaglieri:2017aum}, LDMX~\cite{Berlin:2018bsc,Akesson:2018vlm}, and NA64~($5\times 10^{12}$ electrons on target (EOT))~\cite{Gninenko:2019qiv}. Brown dashed lines correspond to future detectors sensitive to direct scattering signatures of DM particles produced in collider and beam-dump experiments: BDX~\cite{Battaglieri:2017aum}, SND@LHC~\cite{Ahdida:2020evc}, and SND@SHiP~\cite{SHiP:2020noy}. In the case of elastic scalar DM, the additional future reach of the SuperCDMS at SNOLAB and SENSEI DM direct detection experiments~\cite{Battaglieri:2017aum} are also indicated by green dash-dotted lines. }
\label{fig:results}
\end{figure}

In \figref{results}, we present the expected $90\%$ confidence level (CL) exclusion bound in the DM-electron scattering search for the FASER$\nu$2 experiment. We present results for both the Majorana and (inelastic) scalar DM models, assuming $\alpha_D=0.5$ and a fixed mass ratio $m_{A'}=3m_\chi$. As discussed in \secref{model}, these parameter choices are fairly conservative in terms of experimentally testing the thermal DM production hypothesis. As can be seen, in both cases, FASER$\nu$2 will be able to probe the scenarios with $m_\chi\agt \mathcal{O}(10~\mev)$ and the thermal value of the DM relic density that coincides with the Planck observations~\cite{Aghanim:2018eyx}. In the regime $m_A' = 3 m_\chi \agt {\cal O}(10~\mev)$, the DM signal rate scales approximately as $y^2 / m_\chi^4$, while the DM annihilation cross section, \eqref{sigv}, scales as $y/m_\chi^2$, explaining why the FASER$\nu$2 reach line and relic abundance curve have approximately the same slope in~\figref{results}. In contrast, for light dark photons, the total DM-electron scattering cross section and resulting signal rate is largely independent of $m_{A'}$, a feature also observed in~\figref{results}. The spike in the expected FASER$\nu$2 reach for $m_{A^\prime} = 3m_\chi \simeq 770-780~\mev$ is due to the $A^\prime$ mixing with the $\rho$ and $\omega$ mesons, which is taken into account in the dark photon production in proton bremsstrahlung~\cite{Faessler:2009tn,deNiverville:2016rqh}; see also Ref.~\cite{Feng:2017uoz} for a similar discussion for FASER.

In the plots, the dark gray-shaded region corresponds to null searches at BaBar~\cite{Lees:2017lec}, MiniBooNE~\cite{Aguilar-Arevalo:2018wea}, and NA64~\cite{NA64:2019imj}. These accelerator-based constraints are more stringent than those coming from the electron and muon anomalous magnetic moments~\cite{Hanneke:2008tm,Bennett:2006fi}. In addition, in the elastic scalar DM case, bounds from past DM direct detection searches at CRESST-II~\cite{Angloher:2015ewa} and XENON1T~\cite{Aprile:2019xxb} constrain parts of the parameter space. These bounds, however, can be avoided in the inelastic scalar DM model and are, therefore, presented with black lines but no gray-shaded regions. We similarly show the constraints from sensitivity limit recasts of a number of past beam-dump and neutrino experiments including BEBC~\cite{Grassler:1986vr}, CHARM-II~\cite{DeWinter:1989zg}, E137~\cite{Batell:2014mga}, LSND~\cite{deNiverville:2011it}, and NO$\nu$A~\cite{Wang:2017tmk} that are implemented following Ref.~\cite{Marsicano:2018glj,Buonocore:2019esg} and presented with the thin solid black line.

As is clearly illustrated in the plot, the FASER$\nu$2 search for DM scattering events will be complementary to other experimental efforts. In particular, future missing energy and momentum searches at the Belle-II~\cite{Abe:2010gxa} (shown following Ref.~\cite{Battaglieri:2017aum}), NA64~\cite{Gninenko:2019qiv}, and proposed LDMX~\cite{Berlin:2018bsc,Akesson:2018vlm} experiments will be able to independently constrain these scenarios. Crucially, in the case of a discovery, employing such different experimental search strategies will provide important and complementary information about the nature of DM, with FASER$\nu$2 allowing one to directly detect their scattering events and probe the dark coupling constant $\alpha_D$. 
For comparison, we also show the expected sensitivity reach for several proposed experiments sensitive to the scattering of accelerator-produced DM, including the dedicated emulsion detectors of the SND@LHC~\cite{Ahdida:2020evc} and SND@SHiP~\cite{Anelli:2015pba,SHiP:2020noy} experiments, as well as the BDX electron beam dump experiment~\cite{Battaglieri:2016ggd}.

Last, but not least, the elastic scalar DM scenario will be independently probed by future direct detection experiments. We show projections for the SuperCDMS at SNOLAB and SENSEI experiments, following Ref.~\cite{Battaglieri:2017aum}. As can be seen in \figref{results}, these experiments will probe the thermal relic target for a wide range of DM masses with $m_\chi\agt \textrm{a few}~\mev$. In the case of a discovery, FASER$\nu$2 will then provide an independent test of the nature of such DM particles. In contrast, it is much more challenging to directly detect Majorana DM particles, given the suppression of their non-relativistic scattering rates. This is also true for the case of the inelastic scalar DM scenario. These limitations can, however, be easily overcome in the search for accelerator-produced boosted $\chi$s at the LHC in FASER$\nu$2.

\begin{figure}[t]
\begin{center}
\includegraphics*[width=0.49\textwidth]{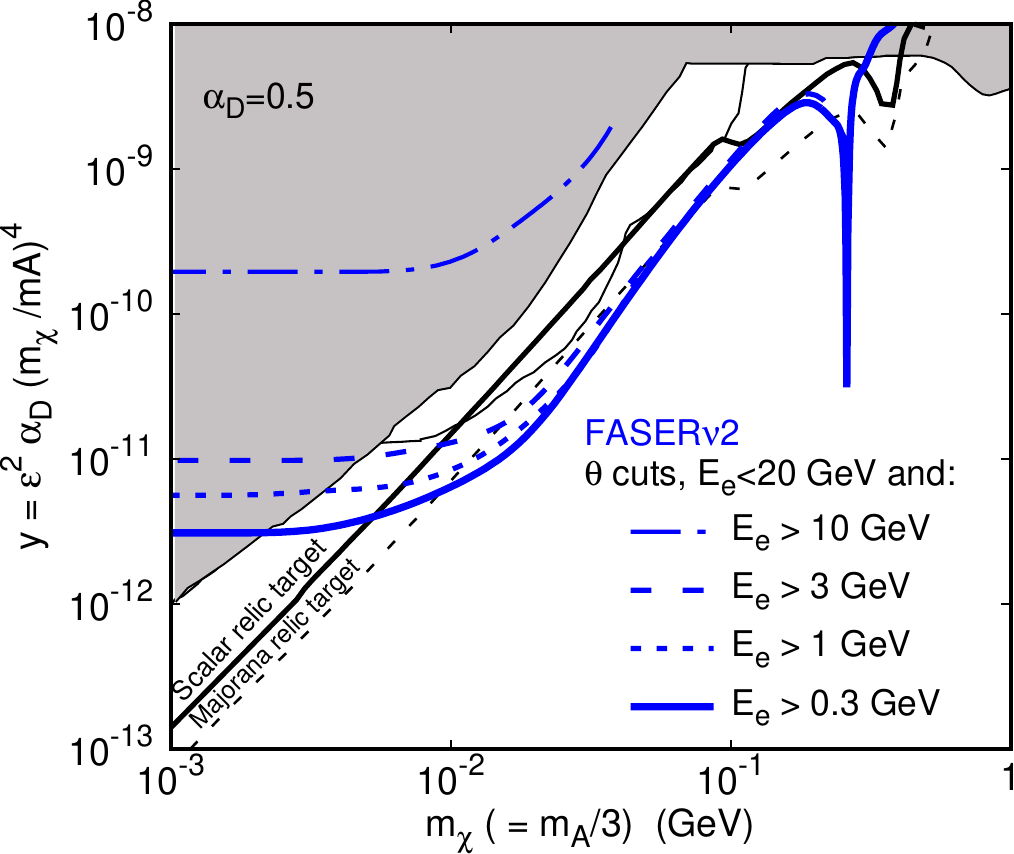}
\hfill
\includegraphics*[width=0.49\textwidth]{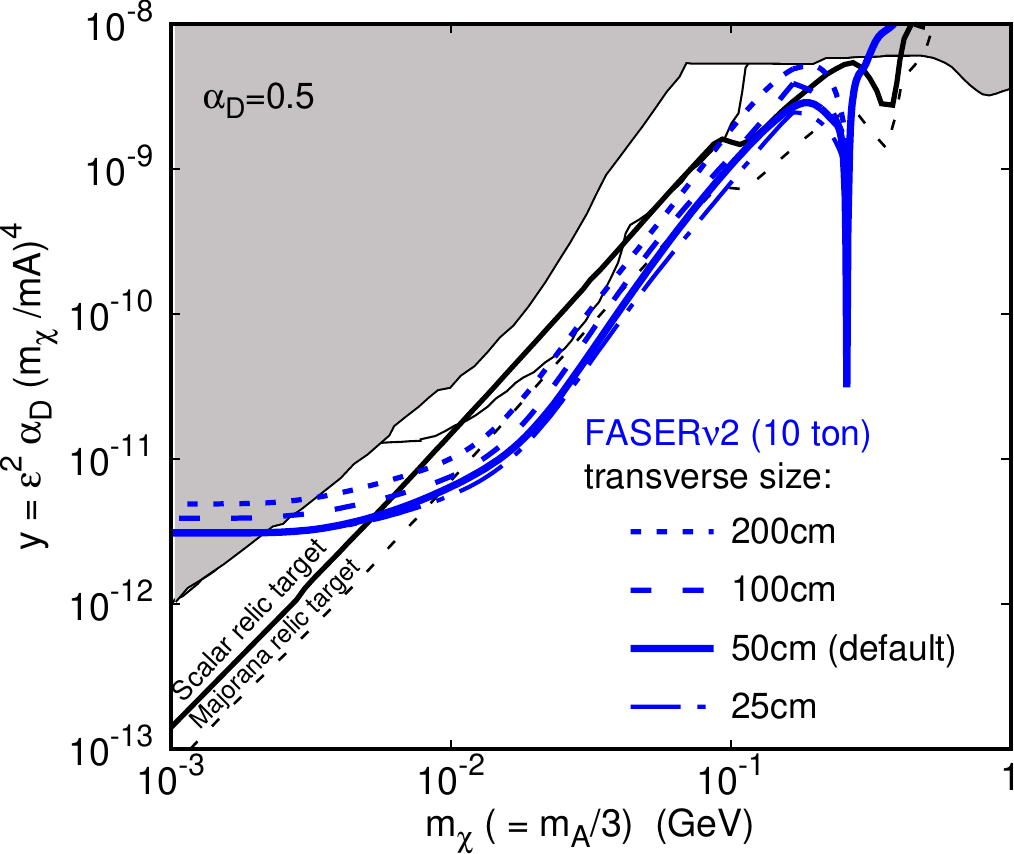}
\end{center}
\vspace*{-0.3in}
\caption{Impact of increasing the electron recoil energy threshold (left) and of varying the detector geometry (right) on the FASER$\nu$2 sensitivity reach. In the left panel, all the cuts are kept the same as in \tableref{cuts} besides the lower limit on $E_e$, which changes as indicated in the plot. In right panel, the total mass (volume) of the $10$-tonne detector is kept constant, while its transverse size and length are changed.}
\label{fig:cutsgeometry}
\end{figure}

In \figref{cutsgeometry} we study the impact of the analysis cuts and of a modified FASER$\nu$2 geometry on the expected sensitivity reach lines. As discussed in \secsref{nubackgrounds}{mubackgrounds}, varying the value of the lower cut on the electron recoil energy does not lead to substantial changes in the neutrino-induced backgrounds, while increasing this value could help to reduce the muon-induced backgrounds. In the left panel of \figref{cutsgeometry}, we show the expected sensitivity reach lines for $E_e>300~\mev$ (default), $1~\gev$, $3~\gev$, and $10~\gev$. In obtaining these results we keep the other cuts the same as shown in \tableref{cuts}. As can be seen, even if the lower limit on $E_e$ is increased to $3~\gev$, FASER$\nu$2 could still constrain important parts of the thermal relic target for both DM scenarios under consideration. On the other hand, if only a very limited range of the electron recoil energies is allowed, $10~\gev<E_e<20~\gev$, then the expected reach is not competitive with current bounds.

In the right panel of \figref{cutsgeometry}, we show the expected reach for a varying FASER$\nu$2 geometry assuming a fixed detector volume. We consider both smaller and larger transverse sizes of the detector $S_T$, as indicated in the plot, while in each case both its length $\Delta$ and the neutrino-induced backgrounds are adjusted and simulated accordingly. As can be seen, a longer, but narrower, detector is preferred to improve the sensitivity reach. Notably, the default geometry of the detector described in \secref{signal} with $S_T=(50~\cm\times 50~\cm)$ transverse size and $\Delta=2~\m$ length appears to be close to optimal. 
The default design may also facilitate a more straightforward detector construction and may more easily fit within the available experimental facility in comparison to the more elongated narrower design with $S_T=(25~\cm\times 25~\cm)$ and $\Delta=8~\m$.

\subsection{Results for \flare} 

\begin{figure}[t]
\begin{center}
\includegraphics*[width=0.49\textwidth]{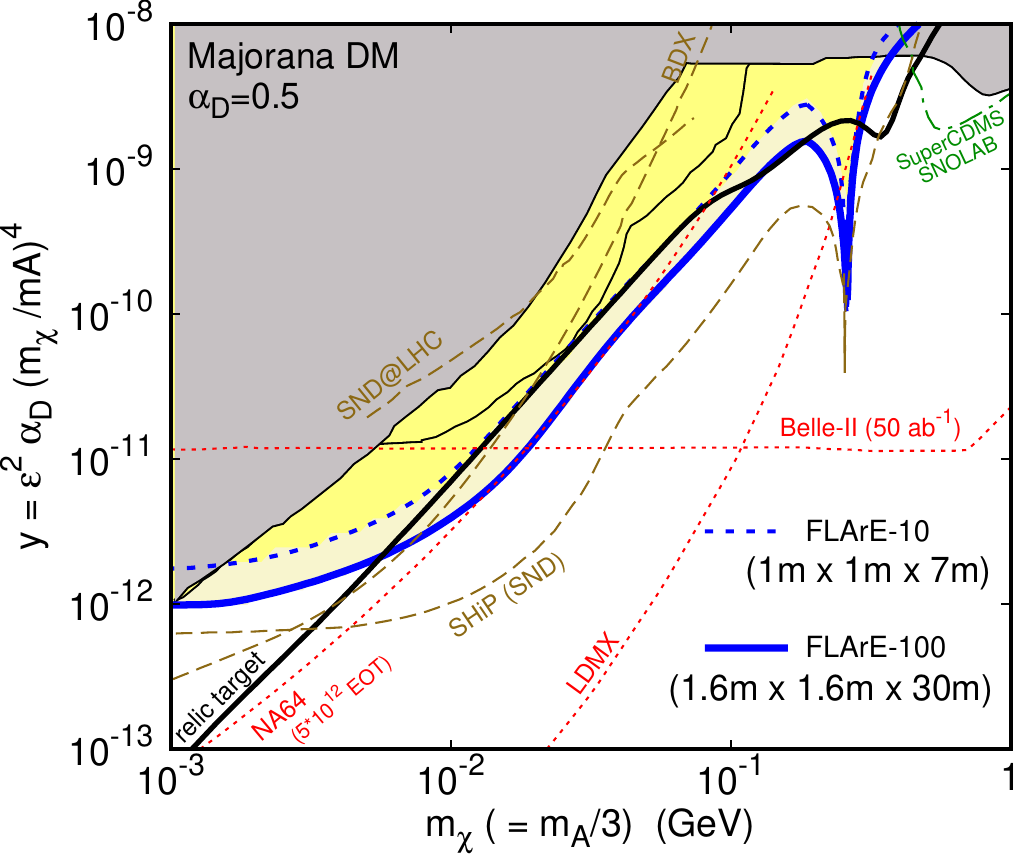}
\hfill
\includegraphics*[width=0.49\textwidth]{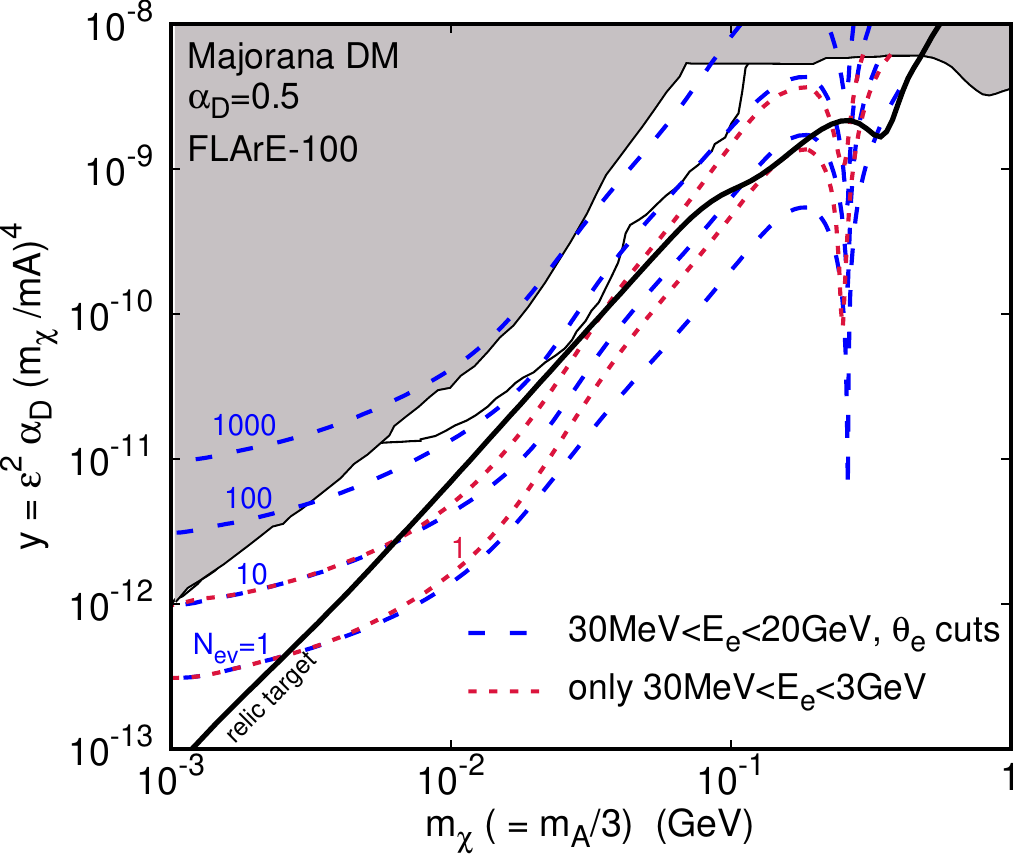}
\end{center}
\vspace*{-0.3in}
\caption{Left: The $90\%$ CL exclusion bounds in the DM-electron scattering search for light Majorana DM for the $10$-tonne liquid-argon detector \flareten (dotted blue line) and the $100$-tonne detector \flarehundred (solid blue) with the designs presented in \secref{Detectors}. The current bounds, DM relic density lines, and future expected sensitivity reach contours for other experiments are shown as in \figref{results}. Right: The dashed blue lines are contours of the number of detected DM scattering events $N_{\textrm{ev}} = 1,10,100,1000$ for \flarehundred and the analysis cuts given in \tableref{cutsLAr}. The red dotted lines correspond to $N_{\textrm{ev}}=1,10$ in \flarehundred, assuming that the analysis employs only the electron recoil energy cut $30~\mev < E_e < 3~\gev$.}
\label{fig:resultsLAr}
\end{figure}

In \figref{resultsLAr}, we present results for the LArTPC detectors.  In the left panel are the expected $90\%$ CL exclusion bounds for the Majorana DM model and both \flareten and \flarehundred. As can be seen, important parts of the relic target line in this model can be probed in the HL-LHC era. The larger detector \flarehundred offers a clear improvement in the expected sensitivity, although the difference between the two benchmark detector designs is not as significant as might be naively expected. Given the 10 times larger volume, one might expect a factor of 3 improvement in the reach in the $y$ parameter of \flarehundred over \flareten, but this gain is somewhat diminished by the larger neutrino-induced backgrounds in \flarehundred.  In addition, the flux of energetic DM particles decreases away from the beam collision axis, and so the increased transverse size of \flarehundred has less of an impact on the expected sensitivity. 

In the right panel of \figref{resultsLAr} we show contours of the expected number of detected DM scattering events in \flarehundred. As can be seen, up to $1000$ DM scattering events are expected during HL-LHC in the region of parameter space not currently excluded by NA64~\cite{NA64:2019imj}. Also shown are the effect of more stringent cuts on the electron recoil energy, $30~\mev < E_e < 3~\gev$. This change in selection cuts with respect to \tableref{cutsLAr} does not significantly degrade the sensitivity reach for light DM particles with $m_\chi\alt 10~\mev$. Such light DM favors a low momentum exchange between $\chi$ and $e^-$ as dictated by the smaller value of the dark photon mass, $m_{A^\prime}=m_{\chi}/3$. On the other hand, for larger DM masses, $m_\chi \agt 100~\mev$, restricting to events with lower electron recoil energy would result in a mild reduction in sensitivity to the $y$ parameter.

We stress that, even though the reach plots presented here look similar for both the emulsion and LArTPC detectors, the final sensitivity will also be affected by the detector capability to reject muon-induced backgrounds. In particular, as discussed in \secref{mubackgrounds}, the LArTPC experiments allow for a very efficient active muon veto and, therefore, present a particularly promising detection technique to be employed in DM searches in the far-forward region of the LHC. Importantly, since the LArTPC technique is by design well-suited to the study of neutrino interactions, such a detector could thus be considered for a future far-forward neutrino experiment during the HL-LHC phase~\cite{SnowmassNeutrinoDetectors}. 

\section{Conclusions}
\label{sec:conclusions}

The hypothesis that DM is part of a light hidden sector is both theoretically compelling and experimentally fertile. We have proposed to leverage the large forward cross section in $pp$ collisions at the LHC to produce and detect such light DM particles in the MeV to GeV mass range. In simple models based on a kinetically-mixed dark photon mediator, an intense flux of DM particles will be produced in the far-forward direction through neutral meson decays or proton bremsstrahlung. Given a suitable detector situated in the forward region, perhaps housed in the proposed Forward Physics Facility~\cite{SnowmassFPF}, the DM particles can then be detected through their elastic scattering with electrons, and as many as hundreds to thousands of such DM-electron scattering events could be detected in the cosmologically-allowed regions of parameter space at the HL-LHC.

This search strategy is complementary to approaches utilizing missing energy/momentum at colliders and electron-fixed target experiments, because the production probes the hadronic couplings of the mediator, and the produced DM is directly detected through its scattering. Furthermore, the relativistic nature of the accelerator-produced DM renders the scattering insensitive to the detailed structure of the DM interactions. This is in contrast to the non-relativistic scattering in direct detection experiments, where in certain models (e.g., inelastic scalar or Majorana fermion DM) the event rates are substantially suppressed.

We have studied two plausible detector designs, one based on the emulsion detector technology, currently used for FASER$\nu$, and another employing the liquid argon time projection chamber concept, which we have named the Forward Liquid Argon Experiment (\flare). Kinematic and topological handles can be utilized to efficiently separate the DM signal from neutrino induced backgrounds, including neutrino-electron elastic scattering and various neutrino charged current reactions. We have also investigated potential background processes induced by the large flux of forward muons. Such backgrounds may pose challenges to the DM search with a nominal FASER$\nu$-style emulsion detector, given the lack of event time information and anticipated spatial pile-up. We have suggested several strategies to mitigate these backgrounds, including a sweeper magnet to deflect the muons, the installation of electronic timing layers in the emulsion detector, and the use of time and spatial information in the LArTPC detector, with the latter approach appearing to be particularly promising. 

Looking ahead, it would be very interesting to investigate the sensitivity of these experiments to the scattering of light DM with nuclei. In contrast to lower-energy, proton beam fixed target experiments, the higher energies of the produced DM particles offer the possibility of detecting DM-induced DIS events. It would also be worthwhile to study the prospects for testing other light DM models with different coupling patterns. For example, the LHC $pp$ collisions may offer distinct advantages in probing DM models with  hadrophilic couplings~\cite{Batell:2014yra,Dobrescu:2014ita,Coloma:2015pih,Batell:2018fqo}.

Although the search for DM particles provides one important motivation for experiments of this kind in the far-forward region of the LHC, it is certainly not the only one. As already highlighted in the introduction, the main goal of FASER$\nu$ and its envisioned HL-LHC successor is to study TeV-energy, collider-produced neutrinos. Along with the large neutrino DIS event rates studied in Refs.~\cite{Abreu:2019yak,Abreu:2020ddv}, we have found there are significant rates for other high-energy neutrino scattering processes, such as neutrino-electron elastic, CCQE, and CCRES scattering. It would be worthwhile to understand the benefits and drawbacks of various detector options, e.g., emulsion and LArTPC, in measuring the various neutrino interactions.  Along these lines, it would be interesting to consider the merits of re-purposing existing detectors in the far-forward region at the LHC, such as the proto-DUNE LArTPC already installed at CERN~\cite{Abi:2017aow,Tsai-Forward-DUNE}. Along with the intrinsic interest in exploring high-energy neutrino interactions, precise measurements of the neutrino flux and a better theoretical description of neutrino scattering in these experiments are of critical importance in the DM search as it relates to the neutrino-induced backgrounds.  Furthermore, besides DM scattering, one can envision other well-motivated BSM scenarios with exotic collider-stable particles that can be detected through their scattering with electrons or nuclei, such as heavy neutral leptons with non standard interactions~\cite{Jodlowski:2020vhr}.

FASER and FASER$\nu$ will soon embark on a groundbreaking physics program exploiting the large forward $pp$ cross section at the LHC. With most of the LHC luminosity still to be collected, now is an apt time to broadly explore the potential physics opportunities afforded by a diverse array of experiments situated in the far-forward region~\cite{SnowmassFPF}. This work demonstrates that, along with the potential for a suite of SM neutrino measurements, a suitable far-forward detector also offers the unique and exciting opportunity to search for light DM at the HL-LHC. 

\acknowledgments

We thank Aki Ariga, Tomoko Ariga, Josh Berger, Jamie Boyd, Milind Diwan, Felix Kling, Marcin Ku\'zniak, Vittorio Paolone, Yu-Dai Tsai, and Masayuki Wada for helpful conversations. We thank Francesco Cerutti and Marta Sabate Gilarte from the CERN STI group for sharing with us the far-forward neutrino spectrum~\cite{Beni:2020yfy} used in our background analysis. The work of BB is supported by the U.S. Department of Energy under grant No. DE--SC0007914. The work of JLF is supported in part by U.S.~National Science Foundation Grant No.~PHY-1915005 and by Simons Investigator Award \#376204. ST is supported by the grant ``AstroCeNT: Particle Astrophysics Science and Technology Centre'' carried out within the International Research Agendas programme of the Foundation for Polish Science financed by the European Union under the European Regional Development Fund. ST is supported in part by the Polish Ministry of Science and Higher Education through its scholarship for young and outstanding scientists (decision no 1190/E-78/STYP/14/2019). At the early stage of the work on this project, ST was also supported by the Lancaster-Manchester-Sheffield  Consortium  for Fundamental Physics under STFC grant ST/P000800/1.

\appendix

\section{Dark Matter-Electron Scattering}
\label{sec:scattering}

Here we provide a few supplementary details regarding the DM-electron elastic scattering cross section. We consider the process
\begin{equation}
\chi(p_1) + e(p_2) \rightarrow \chi(p_3) +e(p_4) \ .
\end{equation}
The kinematics in the lab frame is described by the following four-momenta:
\begin{equation}
p_1 = (E, 0,0,p) \, , ~ p_2 = (m_e,0,0,0)\, , ~
p_3 = (E_\chi, k_\chi s_\chi, 0, k_\chi c_\chi) \, ,~
p_4 = (E_e, k_e s_e, 0, k_e c_e) \, ,~
\end{equation}
where $s_{\chi,e} = \sin \theta_{\chi, e}$ and $c_{\chi,e} = \cos \theta_{\chi, e}$. The 4-momentum transfer is $q = p_1 -p_3 = p_4 - p_2$. 
The Mandelstam variables can be written in terms of the outgoing electron energy as 
\begin{eqnarray}
s & = &  m_\chi^2 + m_e^2 + 2 m_e E\, , \nonumber \\
t & = &  2 m_e^2 - 2 m_e E_e\, ,\nonumber  \\
u & = &  m_\chi^2 - m_e^2 - 2 m_e E + 2 m_e E_e \, .
\label{eq:mandelstam}
\end{eqnarray}

The differential cross section with respect to the outgoing electron energy is 
\begin{equation}
\frac{d \sigma}{dE_e} = \frac{ |\overline{\cal M}|^2 }{32 \pi m_e p^2} \ .
 \label{eq:diffXSdEe}
\end{equation}
In terms of the Mandelstam variables, the squared amplitudes, averaged over initial spins and summed over final spins, for the complex scalar and Majorana DM models are
\begin{align}
\label{eq:squaredamps}
  |\overline{\cal M}|^2 \! = \! \frac{ 16 \pi^2  \epsilon^2  \alpha   \alpha_D  }{ ( t \! - \! m_{A'}^2 )^2 } 
 \! \times \! \begin{cases}
  [ (s-u)^2 + 4 m_\chi^2 t -t^2] ~{\rm (complex~scalar~DM)} \, , \\
  \vspace{-10pt} \\
  2 [s^2 \! + \! u^2 \! - \! 2 m_e^2(s \! - \! t \! + \! u) \! - \! 2 m_\chi^4 \! + \! 2 m_e^4 \! - \!  8 m_\chi^2 m_e^2]~{\rm (Majorana~DM)} \, .
  \end{cases}
\end{align}
Combining Eqs.~(\ref{eq:mandelstam}), (\ref{eq:diffXSdEe}), and (\ref{eq:squaredamps}), we obtain the formulae given in \eqsref{complscat}{Majoscat}.

Distributions in other kinematic variables may be useful in other contexts. For instance, the ``inelasticity'' $y$ is defined as  
\begin{equation}
y = \frac{p_2 \cdot q}{p_2 \cdot p_1 } = \frac{E- E_\chi}{E} = \frac{E_e - m_e}{E} \ ,
\end{equation}
and the differential cross section with respect to the inelasticity is given by
\begin{equation}
\frac{d \sigma}{dy} = \frac{ E \,|\overline{\cal M}|^2 }{32 \pi m_e p^2} \ .
 \label{eq:diffXSdy}
\end{equation}

\bibliography{main}

\providecommand{\href}[2]{#2}\begingroup\raggedright\begin{thebibliography}{100}

\bibitem{Boehm:2003hm}
C.~Boehm and P.~Fayet, ``{Scalar dark matter candidates},''
  \href{http://dx.doi.org/10.1016/j.nuclphysb.2004.01.015}{{\em Nucl. Phys. B}
  {\bfseries 683} (2004) 219--263},
  \href{http://arxiv.org/abs/hep-ph/0305261}{{\ttfamily arXiv:hep-ph/0305261}}.

\bibitem{Pospelov:2007mp}
M.~Pospelov, A.~Ritz, and M.~B. Voloshin, ``{Secluded WIMP Dark Matter},''
  \href{http://dx.doi.org/10.1016/j.physletb.2008.02.052}{{\em Phys. Lett. B}
  {\bfseries 662} (2008) 53--61},
  \href{http://arxiv.org/abs/0711.4866}{{\ttfamily arXiv:0711.4866 [hep-ph]}}.

\bibitem{Feng:2008ya}
J.~L. Feng and J.~Kumar, ``{The WIMPless Miracle: Dark-Matter Particles without
  Weak-Scale Masses or Weak Interactions},''
  \href{http://dx.doi.org/10.1103/PhysRevLett.101.231301}{{\em Phys. Rev.
  Lett.} {\bfseries 101} (2008) 231301},
  \href{http://arxiv.org/abs/0803.4196}{{\ttfamily arXiv:0803.4196 [hep-ph]}}.

\bibitem{Lee:1977ua}
B.~W. Lee and S.~Weinberg, ``{Cosmological Lower Bound on Heavy Neutrino
  Masses},'' \href{http://dx.doi.org/10.1103/PhysRevLett.39.165}{{\em Phys.
  Rev. Lett.} {\bfseries 39} (1977) 165--168}.

\bibitem{Izaguirre:2015yja}
E.~Izaguirre, G.~Krnjaic, P.~Schuster, and N.~Toro, ``{Analyzing the Discovery
  Potential for Light Dark Matter},''
  \href{http://dx.doi.org/10.1103/PhysRevLett.115.251301}{{\em Phys. Rev.
  Lett.} {\bfseries 115} no.~25, (2015) 251301},
  \href{http://arxiv.org/abs/1505.00011}{{\ttfamily arXiv:1505.00011
  [hep-ph]}}.

\bibitem{Krnjaic:2015mbs}
G.~Krnjaic, ``{Probing Light Thermal Dark-Matter With a Higgs Portal
  Mediator},'' \href{http://dx.doi.org/10.1103/PhysRevD.94.073009}{{\em Phys.
  Rev. D} {\bfseries 94} no.~7, (2016) 073009},
  \href{http://arxiv.org/abs/1512.04119}{{\ttfamily arXiv:1512.04119
  [hep-ph]}}.

\bibitem{Batell:2017cmf}
B.~Batell, T.~Han, D.~McKeen, and B.~Shams Es~Haghi, ``{Thermal Dark Matter
  Through the Dirac Neutrino Portal},''
  \href{http://dx.doi.org/10.1103/PhysRevD.97.075016}{{\em Phys. Rev. D}
  {\bfseries 97} no.~7, (2018) 075016},
  \href{http://arxiv.org/abs/1709.07001}{{\ttfamily arXiv:1709.07001
  [hep-ph]}}.

\bibitem{Alexander:2016aln}
J.~Alexander {\em et al.}, ``{Dark Sectors 2016 Workshop: Community Report},''
\newblock 8, 2016.
\newblock \href{http://arxiv.org/abs/1608.08632}{{\ttfamily arXiv:1608.08632
  [hep-ph]}}.

\bibitem{Battaglieri:2017aum}
M.~Battaglieri {\em et al.}, ``{US Cosmic Visions: New Ideas in Dark Matter
  2017: Community Report},'' in {\em {U.S. Cosmic Visions: New Ideas in Dark
  Matter}}.
\newblock 7, 2017.
\newblock \href{http://arxiv.org/abs/1707.04591}{{\ttfamily arXiv:1707.04591
  [hep-ph]}}.

\bibitem{Beacham:2019nyx}
J.~Beacham {\em et al.}, ``{Physics Beyond Colliders at CERN: Beyond the
  Standard Model Working Group Report},''
  \href{http://dx.doi.org/10.1088/1361-6471/ab4cd2}{{\em J. Phys. G} {\bfseries
  47} no.~1, (2020) 010501}, \href{http://arxiv.org/abs/1901.09966}{{\ttfamily
  arXiv:1901.09966 [hep-ex]}}.

\bibitem{Feng:2017uoz}
J.~L. Feng, I.~Galon, F.~Kling, and S.~Trojanowski, ``{ForwArd Search
  ExpeRiment at the LHC},''
  \href{http://dx.doi.org/10.1103/PhysRevD.97.035001}{{\em Phys. Rev.}
  {\bfseries D97} no.~3, (2018) 035001},
\href{http://arxiv.org/abs/1708.09389}{{\ttfamily arXiv:1708.09389 [hep-ph]}}.

\bibitem{Ariga:2018zuc}
{\bfseries FASER} Collaboration, A.~Ariga {\em et al.}, ``{Letter of Intent for
  FASER: ForwArd Search ExpeRiment at the LHC},''
  \href{http://arxiv.org/abs/1811.10243}{{\ttfamily arXiv:1811.10243
  [physics.ins-det]}}.

\bibitem{Ariga:2018pin}
{\bfseries FASER} Collaboration, A.~Ariga {\em et al.}, ``{Technical Proposal
  for FASER: ForwArd Search ExpeRiment at the LHC},''
  \href{http://arxiv.org/abs/1812.09139}{{\ttfamily arXiv:1812.09139
  [physics.ins-det]}}.

\bibitem{Feng:2017vli}
J.~L. Feng, I.~Galon, F.~Kling, and S.~Trojanowski, ``{Dark Higgs bosons at the
  ForwArd Search ExpeRiment},''
  \href{http://dx.doi.org/10.1103/PhysRevD.97.055034}{{\em Phys. Rev. D}
  {\bfseries 97} no.~5, (2018) 055034},
  \href{http://arxiv.org/abs/1710.09387}{{\ttfamily arXiv:1710.09387
  [hep-ph]}}.

\bibitem{Batell:2017kty}
B.~Batell, A.~Freitas, A.~Ismail, and D.~Mckeen, ``{Flavor-specific scalar
  mediators},'' \href{http://dx.doi.org/10.1103/PhysRevD.98.055026}{{\em Phys.
  Rev. D} {\bfseries 98} no.~5, (2018) 055026},
  \href{http://arxiv.org/abs/1712.10022}{{\ttfamily arXiv:1712.10022
  [hep-ph]}}.

\bibitem{Helo:2018qej}
J.~C. Helo, M.~Hirsch, and Z.~S. Wang, ``{Heavy neutral fermions at the
  high-luminosity LHC},'' \href{http://dx.doi.org/10.1007/JHEP07(2018)056}{{\em
  JHEP} {\bfseries 07} (2018) 056},
  \href{http://arxiv.org/abs/1803.02212}{{\ttfamily arXiv:1803.02212
  [hep-ph]}}.

\bibitem{Kling:2018wct}
F.~Kling and S.~Trojanowski, ``{Heavy Neutral Leptons at FASER},''
  \href{http://dx.doi.org/10.1103/PhysRevD.97.095016}{{\em Phys. Rev. D}
  {\bfseries 97} no.~9, (2018) 095016},
  \href{http://arxiv.org/abs/1801.08947}{{\ttfamily arXiv:1801.08947
  [hep-ph]}}.

\bibitem{Feng:2018pew}
J.~L. Feng, I.~Galon, F.~Kling, and S.~Trojanowski, ``{Axionlike particles at
  FASER: The LHC as a photon beam dump},''
  \href{http://dx.doi.org/10.1103/PhysRevD.98.055021}{{\em Phys. Rev. D}
  {\bfseries 98} no.~5, (2018) 055021},
  \href{http://arxiv.org/abs/1806.02348}{{\ttfamily arXiv:1806.02348
  [hep-ph]}}.

\bibitem{Dercks:2018eua}
D.~Dercks, J.~De~Vries, H.~K. Dreiner, and Z.~S. Wang, ``{R-parity Violation
  and Light Neutralinos at CODEX-b, FASER, and MATHUSLA},''
  \href{http://dx.doi.org/10.1103/PhysRevD.99.055039}{{\em Phys. Rev. D}
  {\bfseries 99} no.~5, (2019) 055039},
  \href{http://arxiv.org/abs/1810.03617}{{\ttfamily arXiv:1810.03617
  [hep-ph]}}.

\bibitem{Ariga:2018uku}
{\bfseries FASER} Collaboration, A.~Ariga {\em et al.}, ``{FASER's physics
  reach for long-lived particles},''
  \href{http://dx.doi.org/10.1103/PhysRevD.99.095011}{{\em Phys. Rev. D}
  {\bfseries 99} no.~9, (2019) 095011},
  \href{http://arxiv.org/abs/1811.12522}{{\ttfamily arXiv:1811.12522
  [hep-ph]}}.

\bibitem{Kling:2020mch}
F.~Kling and S.~Trojanowski, ``{Looking forward to test the KOTO anomaly with
  FASER},'' \href{http://dx.doi.org/10.1103/PhysRevD.102.015032}{{\em Phys.
  Rev. D} {\bfseries 102} no.~1, (2020) 015032},
  \href{http://arxiv.org/abs/2006.10630}{{\ttfamily arXiv:2006.10630
  [hep-ph]}}.

\bibitem{Abreu:2019yak}
{\bfseries FASER} Collaboration, H.~Abreu {\em et al.}, ``{Detecting and
  Studying High-Energy Collider Neutrinos with FASER at the LHC},''
  \href{http://dx.doi.org/10.1140/epjc/s10052-020-7631-5}{{\em Eur. Phys. J. C}
  {\bfseries 80} no.~1, (2020) 61},
  \href{http://arxiv.org/abs/1908.02310}{{\ttfamily arXiv:1908.02310
  [hep-ex]}}.

\bibitem{Abreu:2020ddv}
{\bfseries FASER} Collaboration, H.~Abreu {\em et al.}, ``{Technical Proposal:
  FASERnu},'' \href{http://arxiv.org/abs/2001.03073}{{\ttfamily
  arXiv:2001.03073 [physics.ins-det]}}.

\bibitem{Ismail:2020yqc}
A.~Ismail, R.~Mammen~Abraham, and F.~Kling, ``{Neutral current neutrino
  interactions at FASER$\nu$},''
  \href{http://dx.doi.org/10.1103/PhysRevD.103.056014}{{\em Phys. Rev. D}
  {\bfseries 103} no.~5, (2021) 056014},
  \href{http://arxiv.org/abs/2012.10500}{{\ttfamily arXiv:2012.10500
  [hep-ph]}}.

\bibitem{SnowmassFPF}
R.~M. Abraham {\em et al.}, {\em Forward Physics Facility: Snowmass 2021 Letter
  of Interest}, Aug., 2020.
\newblock \url{https://doi.org/10.5281/zenodo.4009640}.

\bibitem{FPFKickoffMeeting}
``{Forward Physics Facility Kickoff Meeting}.''
  \url{https://indico.cern.ch/event/955956}, 2020.

\bibitem{Batell:2009di}
B.~Batell, M.~Pospelov, and A.~Ritz, ``{Exploring Portals to a Hidden Sector
  Through Fixed Targets},''
  \href{http://dx.doi.org/10.1103/PhysRevD.80.095024}{{\em Phys. Rev. D}
  {\bfseries 80} (2009) 095024},
  \href{http://arxiv.org/abs/0906.5614}{{\ttfamily arXiv:0906.5614 [hep-ph]}}.

\bibitem{deNiverville:2011it}
P.~deNiverville, M.~Pospelov, and A.~Ritz, ``{Observing a light dark matter
  beam with neutrino experiments},''
  \href{http://dx.doi.org/10.1103/PhysRevD.84.075020}{{\em Phys. Rev. D}
  {\bfseries 84} (2011) 075020},
  \href{http://arxiv.org/abs/1107.4580}{{\ttfamily arXiv:1107.4580 [hep-ph]}}.

\bibitem{deNiverville:2012ij}
P.~deNiverville, D.~McKeen, and A.~Ritz, ``{Signatures of sub-GeV dark matter
  beams at neutrino experiments},''
  \href{http://dx.doi.org/10.1103/PhysRevD.86.035022}{{\em Phys. Rev. D}
  {\bfseries 86} (2012) 035022},
  \href{http://arxiv.org/abs/1205.3499}{{\ttfamily arXiv:1205.3499 [hep-ph]}}.

\bibitem{Batell:2014yra}
B.~Batell, P.~deNiverville, D.~McKeen, M.~Pospelov, and A.~Ritz, ``{Leptophobic
  Dark Matter at Neutrino Factories},''
  \href{http://dx.doi.org/10.1103/PhysRevD.90.115014}{{\em Phys. Rev. D}
  {\bfseries 90} no.~11, (2014) 115014},
  \href{http://arxiv.org/abs/1405.7049}{{\ttfamily arXiv:1405.7049 [hep-ph]}}.

\bibitem{Dobrescu:2014ita}
B.~A. Dobrescu and C.~Frugiuele, ``{GeV-Scale Dark Matter: Production at the
  Main Injector},'' \href{http://dx.doi.org/10.1007/JHEP02(2015)019}{{\em JHEP}
  {\bfseries 02} (2015) 019}, \href{http://arxiv.org/abs/1410.1566}{{\ttfamily
  arXiv:1410.1566 [hep-ph]}}.

\bibitem{Kahn:2014sra}
Y.~Kahn, G.~Krnjaic, J.~Thaler, and M.~Toups, ``{DAE\ensuremath{\delta}ALUS and
  dark matter detection},''
  \href{http://dx.doi.org/10.1103/PhysRevD.91.055006}{{\em Phys. Rev. D}
  {\bfseries 91} no.~5, (2015) 055006},
  \href{http://arxiv.org/abs/1411.1055}{{\ttfamily arXiv:1411.1055 [hep-ph]}}.

\bibitem{Coloma:2015pih}
P.~Coloma, B.~A. Dobrescu, C.~Frugiuele, and R.~Harnik, ``{Dark matter beams at
  LBNF},'' \href{http://dx.doi.org/10.1007/JHEP04(2016)047}{{\em JHEP}
  {\bfseries 04} (2016) 047}, \href{http://arxiv.org/abs/1512.03852}{{\ttfamily
  arXiv:1512.03852 [hep-ph]}}.

\bibitem{deNiverville:2015mwa}
P.~deNiverville, M.~Pospelov, and A.~Ritz, ``{Light new physics in coherent
  neutrino-nucleus scattering experiments},''
  \href{http://dx.doi.org/10.1103/PhysRevD.92.095005}{{\em Phys. Rev. D}
  {\bfseries 92} no.~9, (2015) 095005},
  \href{http://arxiv.org/abs/1505.07805}{{\ttfamily arXiv:1505.07805
  [hep-ph]}}.

\bibitem{deNiverville:2016rqh}
P.~deNiverville, C.-Y. Chen, M.~Pospelov, and A.~Ritz, ``{Light dark matter in
  neutrino beams: production modelling and scattering signatures at MiniBooNE,
  T2K and SHiP},'' \href{http://dx.doi.org/10.1103/PhysRevD.95.035006}{{\em
  Phys. Rev. D} {\bfseries 95} no.~3, (2017) 035006},
  \href{http://arxiv.org/abs/1609.01770}{{\ttfamily arXiv:1609.01770
  [hep-ph]}}.

\bibitem{DeRomeri:2019kic}
V.~De~Romeri, K.~J. Kelly, and P.~A. Machado, ``{DUNE-PRISM Sensitivity to
  Light Dark Matter},''
  \href{http://dx.doi.org/10.1103/PhysRevD.100.095010}{{\em Phys. Rev. D}
  {\bfseries 100} no.~9, (2019) 095010},
  \href{http://arxiv.org/abs/1903.10505}{{\ttfamily arXiv:1903.10505
  [hep-ph]}}.

\bibitem{Dutta:2019nbn}
B.~Dutta, D.~Kim, S.~Liao, J.-C. Park, S.~Shin, and L.~E. Strigari, ``{Dark
  matter signals from timing spectra at neutrino experiments},''
  \href{http://dx.doi.org/10.1103/PhysRevLett.124.121802}{{\em Phys. Rev.
  Lett.} {\bfseries 124} no.~12, (2020) 121802},
  \href{http://arxiv.org/abs/1906.10745}{{\ttfamily arXiv:1906.10745
  [hep-ph]}}.

\bibitem{Dutta:2020vop}
B.~Dutta, D.~Kim, S.~Liao, J.-C. Park, S.~Shin, L.~E. Strigari, and
  A.~Thompson, ``{Searching for Dark Matter Signals in Timing Spectra at
  Neutrino Experiments},'' \href{http://arxiv.org/abs/2006.09386}{{\ttfamily
  arXiv:2006.09386 [hep-ph]}}.

\bibitem{Aguilar-Arevalo:2017mqx}
{\bfseries MiniBooNE-DM} Collaboration, A.~Aguilar-Arevalo {\em et al.},
  ``{Dark Matter Search in a Proton Beam Dump with MiniBooNE},''
  \href{http://dx.doi.org/10.1103/PhysRevLett.118.221803}{{\em Phys. Rev.
  Lett.} {\bfseries 118} no.~22, (2017) 221803},
  \href{http://arxiv.org/abs/1702.02688}{{\ttfamily arXiv:1702.02688
  [hep-ex]}}.

\bibitem{Aguilar-Arevalo:2018wea}
{\bfseries MiniBooNE-DM} Collaboration, A.~Aguilar-Arevalo {\em et al.},
  ``{Dark Matter Search in Nucleon, Pion, and Electron Channels from a Proton
  Beam Dump with MiniBooNE},''
  \href{http://dx.doi.org/10.1103/PhysRevD.98.112004}{{\em Phys. Rev. D}
  {\bfseries 98} no.~11, (2018) 112004},
  \href{http://arxiv.org/abs/1807.06137}{{\ttfamily arXiv:1807.06137
  [hep-ex]}}.

\bibitem{Holdom:1985ag}
B.~Holdom, ``{Two U(1)'s and Epsilon Charge Shifts},''
  \href{http://dx.doi.org/10.1016/0370-2693(86)91377-8}{{\em Phys. Lett. B}
  {\bfseries 166} (1986) 196--198}.

\bibitem{Ahdida:2020evc}
{\bfseries SHiP} Collaboration, C.~Ahdida {\em et al.}, ``{SND@LHC},''
  \href{http://arxiv.org/abs/2002.08722}{{\ttfamily arXiv:2002.08722
  [physics.ins-det]}}.

\bibitem{Ade:2015xua}
{\bfseries Planck} Collaboration, P.~Ade {\em et al.}, ``{Planck 2015 results.
  XIII. Cosmological parameters},''
  \href{http://dx.doi.org/10.1051/0004-6361/201525830}{{\em Astron. Astrophys.}
  {\bfseries 594} (2016) A13},
  \href{http://arxiv.org/abs/1502.01589}{{\ttfamily arXiv:1502.01589
  [astro-ph.CO]}}.

\bibitem{Slatyer:2009yq}
T.~R. Slatyer, N.~Padmanabhan, and D.~P. Finkbeiner, ``{CMB Constraints on WIMP
  Annihilation: Energy Absorption During the Recombination Epoch},''
  \href{http://dx.doi.org/10.1103/PhysRevD.80.043526}{{\em Phys. Rev. D}
  {\bfseries 80} (2009) 043526},
  \href{http://arxiv.org/abs/0906.1197}{{\ttfamily arXiv:0906.1197
  [astro-ph.CO]}}.

\bibitem{Berlin:2018bsc}
A.~Berlin, N.~Blinov, G.~Krnjaic, P.~Schuster, and N.~Toro, ``{Dark Matter,
  Millicharges, Axion and Scalar Particles, Gauge Bosons, and Other New Physics
  with LDMX},'' \href{http://dx.doi.org/10.1103/PhysRevD.99.075001}{{\em Phys.
  Rev. D} {\bfseries 99} no.~7, (2019) 075001},
  \href{http://arxiv.org/abs/1807.01730}{{\ttfamily arXiv:1807.01730
  [hep-ph]}}.

\bibitem{Feng:2017drg}
J.~L. Feng and J.~Smolinsky, ``{Impact of a resonance on thermal targets for
  invisible dark photon searches},''
  \href{http://dx.doi.org/10.1103/PhysRevD.96.095022}{{\em Phys. Rev. D}
  {\bfseries 96} no.~9, (2017) 095022},
  \href{http://arxiv.org/abs/1707.03835}{{\ttfamily arXiv:1707.03835
  [hep-ph]}}.

\bibitem{Berlin:2020uwy}
A.~Berlin, P.~deNiverville, A.~Ritz, P.~Schuster, and N.~Toro, ``{Sub-GeV dark
  matter production at fixed-target experiments},''
  \href{http://dx.doi.org/10.1103/PhysRevD.102.095011}{{\em Phys. Rev. D}
  {\bfseries 102} no.~9, (2020) 095011},
  \href{http://arxiv.org/abs/2003.03379}{{\ttfamily arXiv:2003.03379
  [hep-ph]}}.

\bibitem{Bernreuther:2020koj}
E.~Bernreuther, S.~Heeba, and F.~Kahlhoefer, ``{Resonant sub-GeV Dirac dark
  matter},'' \href{http://dx.doi.org/10.1088/1475-7516/2021/03/040}{{\em JCAP}
  {\bfseries 03} (2021) 040}, \href{http://arxiv.org/abs/2010.14522}{{\ttfamily
  arXiv:2010.14522 [hep-ph]}}.

\bibitem{Kahn:2020fef}
Y.~Kahn, G.~Krnjaic, and B.~Mandava, ``{Dark Matter Detection With Bound
  Nuclear Targets: The Poisson Phonon Tail},''
  \href{http://arxiv.org/abs/2011.09477}{{\ttfamily arXiv:2011.09477
  [hep-ph]}}.

\bibitem{TuckerSmith:2001hy}
D.~Tucker-Smith and N.~Weiner, ``{Inelastic dark matter},''
  \href{http://dx.doi.org/10.1103/PhysRevD.64.043502}{{\em Phys. Rev. D}
  {\bfseries 64} (2001) 043502},
  \href{http://arxiv.org/abs/hep-ph/0101138}{{\ttfamily arXiv:hep-ph/0101138}}.

\bibitem{Berlin:2018jbm}
A.~Berlin and F.~Kling, ``{Inelastic Dark Matter at the LHC Lifetime Frontier:
  ATLAS, CMS, LHCb, CODEX-b, FASER, and MATHUSLA},''
  \href{http://dx.doi.org/10.1103/PhysRevD.99.015021}{{\em Phys. Rev. D}
  {\bfseries 99} no.~1, (2019) 015021},
  \href{http://arxiv.org/abs/1810.01879}{{\ttfamily arXiv:1810.01879
  [hep-ph]}}.

\bibitem{Jodlowski:2019ycu}
K.~Jod{\l}owski, F.~Kling, L.~Roszkowski, and S.~Trojanowski, ``{Extending the
  reach of FASER, MATHUSLA, and SHiP towards smaller lifetimes using secondary
  particle production},''
  \href{http://dx.doi.org/10.1103/PhysRevD.101.095020}{{\em Phys. Rev. D}
  {\bfseries 101} no.~9, (2020) 095020},
  \href{http://arxiv.org/abs/1911.11346}{{\ttfamily arXiv:1911.11346
  [hep-ph]}}.

\bibitem{SnowmassFASERnu2}
{\bfseries FASER} Collaboration, ``{FASER$\nu$ 2: A Forward Neutrino Experiment
  at the HL LHC}.''
\newblock
  \url{https://www.snowmass21.org/docs/files/summaries/NF/SNOWMASS21-NF10_NF6-EF6_EF9-IF0_FASERnu2-006.pdf}.

\bibitem{Nakamura:2006xs}
T.~Nakamura {\em et al.}, ``{The OPERA film: New nuclear emulsion for
  large-scale, high-precision experiments},''
  \href{http://dx.doi.org/10.1016/j.nima.2005.08.109}{{\em Nucl. Instrum. Meth.
  A} {\bfseries 556} (2006) 80--86}.

\bibitem{Agafonova:2010dc}
{\bfseries OPERA} Collaboration, N.~Agafonova {\em et al.}, ``{Observation of a
  first $\nu_\tau$ candidate in the OPERA experiment in the CNGS beam},''
  \href{http://dx.doi.org/10.1016/j.physletb.2010.06.022}{{\em Phys. Lett. B}
  {\bfseries 691} (2010) 138--145},
  \href{http://arxiv.org/abs/1006.1623}{{\ttfamily arXiv:1006.1623 [hep-ex]}}.

\bibitem{Acquafredda:2009zz}
R.~Acquafredda {\em et al.}, ``{The OPERA experiment in the CERN to Gran Sasso
  neutrino beam},'' \href{http://dx.doi.org/10.1088/1748-0221/4/04/P04018}{{\em
  JINST} {\bfseries 4} (2009) P04018}.

\bibitem{Anelli:2015pba}
{\bfseries SHiP} Collaboration, M.~Anelli {\em et al.}, ``{A facility to Search
  for Hidden Particles (SHiP) at the CERN SPS},''
  \href{http://arxiv.org/abs/1504.04956}{{\ttfamily arXiv:1504.04956
  [physics.ins-det]}}.

\bibitem{Antonello:2015lea}
{\bfseries MicroBooNE, LAr1-ND, ICARUS-WA104} Collaboration, M.~Antonello {\em
  et al.}, ``{A Proposal for a Three Detector Short-Baseline Neutrino
  Oscillation Program in the Fermilab Booster Neutrino Beam},''
  \href{http://arxiv.org/abs/1503.01520}{{\ttfamily arXiv:1503.01520
  [physics.ins-det]}}.

\bibitem{Abi:2020evt}
{\bfseries DUNE} Collaboration, B.~Abi {\em et al.}, ``{Deep Underground
  Neutrino Experiment (DUNE), Far Detector Technical Design Report, Volume II
  DUNE Physics},'' \href{http://arxiv.org/abs/2002.03005}{{\ttfamily
  arXiv:2002.03005 [hep-ex]}}.

\bibitem{Acciarri:2016smi}
{\bfseries MicroBooNE} Collaboration, R.~Acciarri {\em et al.}, ``{Design and
  Construction of the MicroBooNE Detector},''
  \href{http://dx.doi.org/10.1088/1748-0221/12/02/P02017}{{\em JINST}
  {\bfseries 12} no.~02, (2017) P02017},
  \href{http://arxiv.org/abs/1612.05824}{{\ttfamily arXiv:1612.05824
  [physics.ins-det]}}.

\bibitem{MicroBooNE:2019knj}
{\bfseries MicroBooNE} Collaboration, {\em {Progress on a High Sensitivity,
  High Purity, One-Electron-One-Proton Search for the MiniBooNE Low Energy
  Excess in the MicroBooNE Detector}}, 10, 2019.
\newblock \url{https://www.osti.gov/biblio/1573230}.

\bibitem{Milind}
Milind {D}iwan, private communication.

\bibitem{Adams:2019bzt}
{\bfseries MicroBooNE} Collaboration, C.~Adams {\em et al.}, ``{Design and
  construction of the MicroBooNE Cosmic Ray Tagger system},''
  \href{http://dx.doi.org/10.1088/1748-0221/14/04/P04004}{{\em JINST}
  {\bfseries 14} no.~04, (2019) P04004},
  \href{http://arxiv.org/abs/1901.02862}{{\ttfamily arXiv:1901.02862
  [physics.ins-det]}}.

\bibitem{CRMC}
C.~Baus, T.~Pierog, and R.~Ulrich, ``{Cosmic Ray Monte Carlo (CRMC)},''.
  \url{https://web.ikp.kit.edu/rulrich/crmc.html}.

\bibitem{Pierog:2013ria}
T.~Pierog, I.~Karpenko, J.~M. Katzy, E.~Yatsenko, and K.~Werner, ``{EPOS LHC:
  Test of collective hadronization with data measured at the CERN Large Hadron
  Collider},'' \href{http://dx.doi.org/10.1103/PhysRevC.92.034906}{{\em Phys.
  Rev.} {\bfseries C92} (2015) 034906},
\href{http://arxiv.org/abs/1306.0121}{{\ttfamily arXiv:1306.0121 [hep-ph]}}.

\bibitem{Sjostrand:2014zea}
T.~Sj\"ostrand, S.~Ask, J.~R. Christiansen, R.~Corke, N.~Desai, P.~Ilten,
  S.~Mrenna, S.~Prestel, C.~O. Rasmussen, and P.~Z. Skands, ``{An introduction
  to PYTHIA 8.2},'' \href{http://dx.doi.org/10.1016/j.cpc.2015.01.024}{{\em
  Comput. Phys. Commun.} {\bfseries 191} (2015) 159--177},
  \href{http://arxiv.org/abs/1410.3012}{{\ttfamily arXiv:1410.3012 [hep-ph]}}.

\bibitem{Blumlein:2013cua}
J.~Bl\"umlein and J.~Brunner, ``{New Exclusion Limits on Dark Gauge Forces from
  Proton Bremsstrahlung in Beam-Dump Data},''
  \href{http://dx.doi.org/10.1016/j.physletb.2014.02.029}{{\em Phys. Lett. B}
  {\bfseries 731} (2014) 320--326},
  \href{http://arxiv.org/abs/1311.3870}{{\ttfamily arXiv:1311.3870 [hep-ph]}}.

\bibitem{Marsicano:2018glj}
L.~Marsicano, M.~Battaglieri, M.~Bond\'\i{}, C.~R. Carvajal, A.~Celentano,
  M.~De~Napoli, R.~De~Vita, E.~Nardi, M.~Raggi, and P.~Valente, ``{Novel Way to
  Search for Light Dark Matter in Lepton Beam-Dump Experiments},''
  \href{http://dx.doi.org/10.1103/PhysRevLett.121.041802}{{\em Phys. Rev.
  Lett.} {\bfseries 121} no.~4, (2018) 041802},
  \href{http://arxiv.org/abs/1807.05884}{{\ttfamily arXiv:1807.05884
  [hep-ex]}}.

\bibitem{Celentano:2020vtu}
A.~Celentano, L.~Darm\'e, L.~Marsicano, and E.~Nardi, ``{New production
  channels for light dark matter in hadronic showers},''
  \href{http://dx.doi.org/10.1103/PhysRevD.102.075026}{{\em Phys. Rev. D}
  {\bfseries 102} no.~7, (2020) 075026},
  \href{http://arxiv.org/abs/2006.09419}{{\ttfamily arXiv:2006.09419
  [hep-ph]}}.

\bibitem{Acciarri:2015uup}
{\bfseries DUNE} Collaboration, R.~Acciarri {\em et al.}, ``{Long-Baseline
  Neutrino Facility (LBNF) and Deep Underground Neutrino Experiment (DUNE)}:
  {Conceptual Design Report, Volume 2: The Physics Program for DUNE at LBNF},''
  \href{http://arxiv.org/abs/1512.06148}{{\ttfamily arXiv:1512.06148
  [physics.ins-det]}}.

\bibitem{Ballett:2016opr}
P.~Ballett, S.~Pascoli, and M.~Ross-Lonergan, ``{MeV-scale sterile neutrino
  decays at the Fermilab Short-Baseline Neutrino program},''
  \href{http://dx.doi.org/10.1007/JHEP04(2017)102}{{\em JHEP} {\bfseries 04}
  (2017) 102}, \href{http://arxiv.org/abs/1610.08512}{{\ttfamily
  arXiv:1610.08512 [hep-ph]}}.

\bibitem{Batell:2019nwo}
B.~Batell, J.~Berger, and A.~Ismail, ``{Probing the Higgs Portal at the
  Fermilab Short-Baseline Neutrino Experiments},''
  \href{http://dx.doi.org/10.1103/PhysRevD.100.115039}{{\em Phys. Rev. D}
  {\bfseries 100} no.~11, (2019) 115039},
  \href{http://arxiv.org/abs/1909.11670}{{\ttfamily arXiv:1909.11670
  [hep-ph]}}.

\bibitem{Chen:2007ae}
{\bfseries MicroBooNE} Collaboration, H.~Chen {\em et al.}, {\em {Proposal for
  a New Experiment Using the Booster and NuMI Neutrino Beamlines: MicroBooNE}},
  10, 2007.
\newblock
  \url{https://lss.fnal.gov/archive/test-proposal/0000/fermilab-proposal-0974.pdf}.

\bibitem{Formaggio:2013kya}
J.~Formaggio and G.~Zeller, ``{From eV to EeV: Neutrino Cross Sections Across
  Energy Scales},'' \href{http://dx.doi.org/10.1103/RevModPhys.84.1307}{{\em
  Rev. Mod. Phys.} {\bfseries 84} (2012) 1307--1341},
  \href{http://arxiv.org/abs/1305.7513}{{\ttfamily arXiv:1305.7513 [hep-ex]}}.

\bibitem{Andreopoulos:2009rq}
C.~Andreopoulos {\em et al.}, ``{The GENIE Neutrino Monte Carlo Generator},''
  \href{http://dx.doi.org/10.1016/j.nima.2009.12.009}{{\em Nucl. Instrum. Meth.
  A} {\bfseries 614} (2010) 87--104},
  \href{http://arxiv.org/abs/0905.2517}{{\ttfamily arXiv:0905.2517 [hep-ph]}}.

\bibitem{Andreopoulos:2015wxa}
C.~Andreopoulos, C.~Barry, S.~Dytman, H.~Gallagher, T.~Golan, R.~Hatcher,
  G.~Perdue, and J.~Yarba, ``{The GENIE Neutrino Monte Carlo Generator: Physics
  and User Manual},'' \href{http://arxiv.org/abs/1510.05494}{{\ttfamily
  arXiv:1510.05494 [hep-ph]}}.

\bibitem{Beni:2020yfy}
N.~Beni {\em et al.}, ``{Further studies on the physics potential of an
  experiment using LHC neutrinos},''
  \href{http://dx.doi.org/10.1088/1361-6471/aba7ad}{{\em J. Phys. G} {\bfseries
  47} no.~12, (2020) 125004}, \href{http://arxiv.org/abs/2004.07828}{{\ttfamily
  arXiv:2004.07828 [hep-ex]}}.

\bibitem{Ferrari:2005zk}
A.~Ferrari, P.~R. Sala, A.~Fasso, and J.~Ranft, ``{FLUKA: A multi-particle
  transport code (Program version 2005)},''.
  \url{https://www.osti.gov/biblio/877507-sC9S9L}.

\bibitem{Battistoni:2015epi}
G.~Battistoni {\em et al.}, ``{Overview of the FLUKA code},''
  \href{http://dx.doi.org/10.1016/j.anucene.2014.11.007}{{\em Annals Nucl.
  Energy} {\bfseries 82} (2015) 10--18}.

\bibitem{FLUKAstudy}
{CERN Sources, Targets, and Interactions Group}, M.~Sabate-Gilarte, F.~Cerutti,
  and A.~Tsinganis, ``Characterization of the radiation field for the FASER
  experiment,'' tech. rep., 2018.

\bibitem{Bai:2020ukz}
W.~Bai, M.~Diwan, M.~V. Garzelli, Y.~S. Jeong, and M.~H. Reno, ``{Far-forward
  neutrinos at the Large Hadron Collider},''
  \href{http://dx.doi.org/10.1007/JHEP06(2020)032}{{\em JHEP} {\bfseries 06}
  (2020) 032}, \href{http://arxiv.org/abs/2002.03012}{{\ttfamily
  arXiv:2002.03012 [hep-ph]}}.

\bibitem{Zyla:2020zbs}
{\bfseries Particle Data Group} Collaboration, P.~Zyla {\em et al.}, ``{Review
  of Particle Physics},'' \href{http://dx.doi.org/10.1093/ptep/ptaa104}{{\em
  PTEP} {\bfseries 2020} no.~8, (2020) 083C01}.

\bibitem{Groom:2001kq}
D.~E. Groom, N.~V. Mokhov, and S.~I. Striganov, ``{Muon stopping power and
  range tables 10-MeV to 100-TeV},''
  \href{http://dx.doi.org/10.1006/adnd.2001.0861}{{\em Atom. Data Nucl. Data
  Tabl.} {\bfseries 78} (2001) 183--356}.

\bibitem{Juget:2009zz}
F.~Juget, ``{Electromagnetic shower reconstruction with emulsion films in the
  OPERA experiment},''
  \href{http://dx.doi.org/10.1088/1742-6596/160/1/012033}{{\em J. Phys. Conf.
  Ser.} {\bfseries 160} (2009) 012033}.

\bibitem{Aghanim:2018eyx}
{\bfseries Planck} Collaboration, N.~Aghanim {\em et al.}, ``{Planck 2018
  results. VI. Cosmological parameters},''
  \href{http://dx.doi.org/10.1051/0004-6361/201833910}{{\em Astron. Astrophys.}
  {\bfseries 641} (2020) A6}, \href{http://arxiv.org/abs/1807.06209}{{\ttfamily
  arXiv:1807.06209 [astro-ph.CO]}}.

\bibitem{Faessler:2009tn}
A.~Faessler, M.~Krivoruchenko, and B.~Martemyanov, ``{Once more on
  electromagnetic form factors of nucleons in extended vector meson dominance
  model},'' \href{http://dx.doi.org/10.1103/PhysRevC.82.038201}{{\em Phys. Rev.
  C} {\bfseries 82} (2010) 038201},
  \href{http://arxiv.org/abs/0910.5589}{{\ttfamily arXiv:0910.5589 [hep-ph]}}.

\bibitem{Akesson:2018vlm}
{\bfseries LDMX} Collaboration, T.~\r{A}kesson {\em et al.}, ``{Light Dark
  Matter eXperiment (LDMX)},''
  \href{http://arxiv.org/abs/1808.05219}{{\ttfamily arXiv:1808.05219
  [hep-ex]}}.

\bibitem{Gninenko:2019qiv}
S.~Gninenko, D.~Kirpichnikov, M.~Kirsanov, and N.~Krasnikov, ``{Combined search
  for light dark matter with electron and muon beams at NA64},''
  \href{http://dx.doi.org/10.1016/j.physletb.2019.07.015}{{\em Phys. Lett. B}
  {\bfseries 796} (2019) 117--122},
  \href{http://arxiv.org/abs/1903.07899}{{\ttfamily arXiv:1903.07899
  [hep-ph]}}.

\bibitem{SHiP:2020noy}
{\bfseries SHiP} Collaboration, ``{Sensitivity of the SHiP experiment to light
  dark matter},'' \href{http://arxiv.org/abs/2010.11057}{{\ttfamily
  arXiv:2010.11057 [hep-ex]}}.

\bibitem{Lees:2017lec}
{\bfseries BaBar} Collaboration, J.~Lees {\em et al.}, ``{Search for Invisible
  Decays of a Dark Photon Produced in ${e}^{+}{e}^{-}$ Collisions at BaBar},''
  \href{http://dx.doi.org/10.1103/PhysRevLett.119.131804}{{\em Phys. Rev.
  Lett.} {\bfseries 119} no.~13, (2017) 131804},
  \href{http://arxiv.org/abs/1702.03327}{{\ttfamily arXiv:1702.03327
  [hep-ex]}}.

\bibitem{NA64:2019imj}
D.~Banerjee {\em et al.}, ``{Dark matter search in missing energy events with
  NA64},'' \href{http://dx.doi.org/10.1103/PhysRevLett.123.121801}{{\em Phys.
  Rev. Lett.} {\bfseries 123} no.~12, (2019) 121801},
  \href{http://arxiv.org/abs/1906.00176}{{\ttfamily arXiv:1906.00176
  [hep-ex]}}.

\bibitem{Hanneke:2008tm}
D.~Hanneke, S.~Fogwell, and G.~Gabrielse, ``{New Measurement of the Electron
  Magnetic Moment and the Fine Structure Constant},''
  \href{http://dx.doi.org/10.1103/PhysRevLett.100.120801}{{\em Phys. Rev.
  Lett.} {\bfseries 100} (2008) 120801},
  \href{http://arxiv.org/abs/0801.1134}{{\ttfamily arXiv:0801.1134
  [physics.atom-ph]}}.

\bibitem{Bennett:2006fi}
{\bfseries Muon g-2} Collaboration, G.~Bennett {\em et al.}, ``{Final Report of
  the Muon E821 Anomalous Magnetic Moment Measurement at BNL},''
  \href{http://dx.doi.org/10.1103/PhysRevD.73.072003}{{\em Phys. Rev. D}
  {\bfseries 73} (2006) 072003},
  \href{http://arxiv.org/abs/hep-ex/0602035}{{\ttfamily arXiv:hep-ex/0602035}}.

\bibitem{Angloher:2015ewa}
{\bfseries CRESST} Collaboration, G.~Angloher {\em et al.}, ``{Results on light
  dark matter particles with a low-threshold CRESST-II detector},''
  \href{http://dx.doi.org/10.1140/epjc/s10052-016-3877-3}{{\em Eur. Phys. J. C}
  {\bfseries 76} no.~1, (2016) 25},
  \href{http://arxiv.org/abs/1509.01515}{{\ttfamily arXiv:1509.01515
  [astro-ph.CO]}}.

\bibitem{Aprile:2019xxb}
{\bfseries XENON} Collaboration, E.~Aprile {\em et al.}, ``{Light Dark Matter
  Search with Ionization Signals in XENON1T},''
  \href{http://dx.doi.org/10.1103/PhysRevLett.123.251801}{{\em Phys. Rev.
  Lett.} {\bfseries 123} no.~25, (2019) 251801},
  \href{http://arxiv.org/abs/1907.11485}{{\ttfamily arXiv:1907.11485
  [hep-ex]}}.

\bibitem{Grassler:1986vr}
{\bfseries BEBC WA66} Collaboration, H.~Grassler {\em et al.}, ``{Prompt
  Neutrino Production in 400-\{GeV\} Proton Copper Interactions},''
  \href{http://dx.doi.org/10.1016/0550-3213(86)90246-4}{{\em Nucl. Phys. B}
  {\bfseries 273} (1986) 253--274}.

\bibitem{DeWinter:1989zg}
{\bfseries CHARM-II} Collaboration, K.~De~Winter {\em et al.}, ``{A Detector
  for the Study of Neutrino - Electron Scattering},''
  \href{http://dx.doi.org/10.1016/0168-9002(89)91190-X}{{\em Nucl. Instrum.
  Meth. A} {\bfseries 278} (1989) 670}.

\bibitem{Batell:2014mga}
B.~Batell, R.~Essig, and Z.~Surujon, ``{Strong Constraints on Sub-GeV Dark
  Sectors from SLAC Beam Dump E137},''
  \href{http://dx.doi.org/10.1103/PhysRevLett.113.171802}{{\em Phys. Rev.
  Lett.} {\bfseries 113} no.~17, (2014) 171802},
  \href{http://arxiv.org/abs/1406.2698}{{\ttfamily arXiv:1406.2698 [hep-ph]}}.

\bibitem{Wang:2017tmk}
{\bfseries NOvA} Collaboration, B.~Wang, J.~Bian, T.~E. Coan, S.~Kotelnikov,
  H.~Duyang, and A.~Hatzikoutelis, ``{Muon Neutrino on Electron Elastic
  Scattering in the NOvA Near Detector and its Applications Beyond the Standard
  Model},'' \href{http://dx.doi.org/10.1088/1742-6596/888/1/012123}{{\em J.
  Phys. Conf. Ser.} {\bfseries 888} no.~1, (2017) 012123}.

\bibitem{Buonocore:2019esg}
L.~Buonocore, C.~Frugiuele, and P.~deNiverville, ``{Hunt for sub-GeV dark
  matter at neutrino facilities: A survey of past and present experiments},''
  \href{http://dx.doi.org/10.1103/PhysRevD.102.035006}{{\em Phys. Rev. D}
  {\bfseries 102} no.~3, (2020) 035006},
  \href{http://arxiv.org/abs/1912.09346}{{\ttfamily arXiv:1912.09346
  [hep-ph]}}.

\bibitem{Abe:2010gxa}
{\bfseries Belle-II} Collaboration, T.~Abe {\em et al.}, ``{Belle II Technical
  Design Report},'' \href{http://arxiv.org/abs/1011.0352}{{\ttfamily
  arXiv:1011.0352 [physics.ins-det]}}.

\bibitem{Battaglieri:2016ggd}
{\bfseries BDX} Collaboration, M.~Battaglieri {\em et al.}, ``{Dark Matter
  Search in a Beam-Dump eXperiment (BDX) at Jefferson Lab},''
  \href{http://arxiv.org/abs/1607.01390}{{\ttfamily arXiv:1607.01390
  [hep-ex]}}.

\bibitem{SnowmassNeutrinoDetectors}
H.~Abreu {\em et al.}, ``{Neutrino / Dark Particle Detectors for the HL-LHC
  Forward Beam}.''
\newblock
  \url{https://www.snowmass21.org/docs/files/summaries/NF/SNOWMASS21-NF10_NF0-EF0_EF0_Ariga-072.pdf}.

\bibitem{Batell:2018fqo}
B.~Batell, A.~Freitas, A.~Ismail, and D.~Mckeen, ``{Probing Light Dark Matter
  with a Hadrophilic Scalar Mediator},''
  \href{http://dx.doi.org/10.1103/PhysRevD.100.095020}{{\em Phys. Rev. D}
  {\bfseries 100} no.~9, (2019) 095020},
  \href{http://arxiv.org/abs/1812.05103}{{\ttfamily arXiv:1812.05103
  [hep-ph]}}.

\bibitem{Abi:2017aow}
{\bfseries DUNE} Collaboration, B.~Abi {\em et al.}, ``{The Single-Phase
  ProtoDUNE Technical Design Report},''
  \href{http://arxiv.org/abs/1706.07081}{{\ttfamily arXiv:1706.07081
  [physics.ins-det]}}.

\bibitem{Tsai-Forward-DUNE}
Y.-D. Tsai, ``{FORMOSA \& Forward-DUNE - Looking Forward to Millicharged Dark
  Sectors and New Neutrino Physics},'' {2020}.
\newblock
  \url{https://indico.cern.ch/event/955956/contributions/4085706/attachments/2139585/3604771/FORMOSA_Forward-DUNE_Tsai.pdf}.
  {Forward Physics Facility - Kickoff Meeting}.

\bibitem{Jodlowski:2020vhr}
K.~Jod\l{}owski and S.~Trojanowski, ``{Neutrino beam-dump experiment with FASER
  at the LHC},'' \href{http://arxiv.org/abs/2011.04751}{{\ttfamily
  arXiv:2011.04751 [hep-ph]}}.

\end{thebibliography}\endgroup

\end{document}